\documentclass[12pt,english,letterpaper]{paper}
\usepackage[T1]{fontenc}
\usepackage[latin9]{inputenc}
\usepackage{float}
\usepackage{amsthm}
\usepackage{amsmath}
\usepackage{graphicx}

\makeatletter

\providecommand{\tabularnewline}{\\}

\theoremstyle{plain}
\newcommand{\lyxaddress}[1]{
\par {\raggedright #1
\vspace{1.4em}
\noindent\par}
}

\usepackage{times}

\usepackage{amsfonts}

\usepackage{amsthm}



\makeatother

\usepackage{babel}

\begin{document}

\title{Efficient equilibrium sampling of all-atom peptides using library-based
Monte Carlo }

\author{Ying Ding, Artem B. Mamonov and Daniel M. Zuckerman%
\thanks{Electronic mail: ddmmzz@pitt.edu%
}}

\maketitle

\lyxaddress{\begin{center}
Department of Computational Biology, School of Medicine, University
of Pittsburgh, Pittsburgh, Pennsylvania 15260
\par\end{center}}
\begin{abstract}
We applied our previously developed library-based Monte Carlo (LBMC)
to equilibrium sampling of several implicitly solvated all-atom peptides.
LBMC can perform equilibrium sampling of molecules using the pre-calculated
statistical libraries of molecular-fragment configurations and energies.
For this study, we employed residue-based fragments distributed according
to the Boltzmann factor of the OPLS-AA forcefield describing the individual
fragments. Two solvent models were employed: a simple uniform dielectric
and the Generalized Born/Surface Area (GBSA) model. The efficiency
of LBMC was compared to standard Langevin dynamics (LD) using three
different statistical tools. The statistical analyses indicate that
LBMC is more than 100 times faster than LD not only for the simple
solvent model but also for GBSA.
\end{abstract}

\section{Introduction}

Conformational fluctuations in proteins are well appreciated to be
essential to biochemistry, in roles accompanying binding, catalysis
and locomotion \cite{Jeremy2006}. In recent years, the importance
of fluctuations has been further underscored by recognition of the
widespread presence of disordered regions in proteins \cite{Dunker2001,Dunker2002}.
Structural experiments, however, are fairly limited in their power
to characterize such fluctuations from a true ensemble perspective.
For a given protein, X-ray crystallography generates one or a very
small number of configurations, typically excluding the most flexible
regions \cite{DePristo2004,Eyal2005}. NMR studies yield highly approximate
structure sets based on simplified forcefields and non-canonical algorithms
\cite{Spronk2003}. Cryo-EM can characterize large structural fluctuations
but at low resolution \cite{Saibil2000}.

In principle, computations are ideal for characterizing fluctuations
in biomolecules, but sampling power is typically inadequate except
for small systems. The basic reason is the well-known gap in timescales
between simulation and biological behavior \cite{Freddolino2008}.
To bridge this gap, much effort has been undertaken in the field to
develop new efficient sampling techniques. Many of these techniques
are based on {}``generalized ensembles'' including replica exchange
\cite{Swendsen1986,Geyer1991,Berg1992,Okamoto2004}. Other techniques
use modified energy surfaces \cite{Iftimie2000,Gelb2003,Hetenyi2002}
or modified dynamics \cite{Zhu2002,Rosso2002,Minary2008,Abrams2008,Maragliano2006}.
The {}``resolution exchange'' (ResEx) algorithm uses a ladder of
different resolution models with occasional exchanges between levels
\cite{Lyman2006a}. Both replica-exchange and ResEx can be implemented
in the serial top-down scheme \cite{Frantz1990,Lyman2006b}.

Another strategy for speeding calculations is to exploit computer
memory to store frequently used information. In particular, libraries
of molecular fragment configurations can be stored and re-used. Libraries
were previous used by Rosetta \cite{Rohl2004a}, but not for true
canonical sampling. In our previous work we introduced the statistical
rigorous library based Monte Carlo (LBMC) and used it to incorporate
atomic details into a coarse-grained protein model at a small computational
cost \cite{Mamonov2009}. The semi-atomistic model was applied to
equilibrium sampling of several proteins containing up to 309 residues.
LBMC was also applied to equilibrium sampling of several peptides
described by OPLS-AA forcefield \cite{Jorgensen1996} with a simple
uniform dielectric model to model the solvent \cite{Mamonov2009}.
A large efficiency gain of LBMC compared to standard Langevin dynamics
was observed. Inspired by the results of our previous study , here
we further investigate the application of LBMC to equilibrium sampling
of all-atom peptides.

In this study we apply LBMC to several implicitly solvated peptides
described by OPLS-AA forcefield with two different implicit solvent
models: a simple uniform dielectric of 60 and the Generalized Born/Surface
Area (GBSA) model \cite{Qiu1997}. The efficiency of LBMC was quantified
by comparison to Langevin Dynamics using three different statistical
tools. The first tool is based on the autocorrelation behavior of
the end-to-end distance, the second uses our previously developed
{}``decorrelation time'' analysis \cite{Lyman2007} , and the third
employs a block averaging analysis \cite{Flyvbjerg1989} of the end-to-end
distances. All the analyses point to efficiency gains of two to three
orders of magnitude in the three peptides studied - tetraalanine,
octaalanine and Met-enkephalin.

\section{Methods}

\subsection{Library-based Monte Carlo Method (LBMC)}

The library-based Monte Carlo (LBMC) method was recently introduced,
along with complete derivations \cite{Mamonov2009}. Here we briefly
review the method. 

LBMC uses the simple idea to divide a molecule into non-overlapping
fragments, each of which is pre-sampled into a library of Boltzmann-distributed
fragment configurations. For peptides and proteins, fragments based
on amino acids are natural. Trial moves consist of swapping the present
configuration of one or more fragments with members of the corresponding
libraries. LBMC, which is a rigorous MC scheme , has several noteworthy
features. (i) Libraries \textendash{} e.g., for each amino acid \textendash{}
are generated one time and can be re-used in multiple simulations;
accordingly, the internal-to-fragment interactions are never calculated
during a simulation, saving some CPU cost. (ii) Because fragment configurations
are pre-sampled based on all interactions internal to the fragment,
those energy terms do not enter the Metropolis acceptance criterion.
(iii) Perhaps most importantly, the complex correlations among degrees
of freedom internal to a fragment are fully accounted for in the library-generation
stage \textendash{} i.e. , the \textquotedblleft{}price\textquotedblright{}
for the internal timescales is paid in advance. 

A Metropolis-Hastings criterion for an LBMC trial move is derived
in the usual way based on the detailed-balance condition \cite{Mamonov2009}.
In outline, the derivation is accomplished by first separating the
full set of degrees of freedom $\vec{r}$ into $M$ fragments,$\overrightarrow{r}=\{\vec{r}_{1},...,\vec{r}_{M}\}$
. Similarly, the total energy of a forcefield $U^{\mathrm{{tot}}}$,
which could include implicit solvent terms, is decomposed into components:
the terms internal to each fragment $i$, denoted $U_{i}^{\mathrm{{frag}}}$,
and all the \textquotedblleft{}rest,\textquotedblright{} which are
cross-fragment interaction terms lumped into $U^{\mathrm{{rest}}}$.
Thus we have 

\begin{equation}
U^{\mathrm{{tot}}}(\vec{r}_{1},...,\vec{r}_{M})=\sum U_{i}^{\mathrm{{frag}}}(\vec{r}_{i})+U^{\mathrm{{rest}}}(\vec{r}_{1},...,\vec{r}_{M})\label{eq:energy decomposition}\end{equation}

\begin{flushleft}
In the present study, $U^{\mathrm{tot}}$ represent the OPLS-AA forcefield
plus an implicit solvent model, as described below.
\par\end{flushleft}

In our previous work, we derived Metropolis criteria for two types
of library-swap trial moves \cite{Mamonov2009}. The first is a simple
swap move in which configurations from one or more fragments are swapped
with configurations chosen uniformly from the corresponding libraries
(Each library is already Boltzmann distributed according to $U_{i}^{\mathrm{{frag}}}$,
as described below.) In the simple swap move, the generating probability
for the trial/new configuration ($n$) is completely independent of
the old configuration ($o$). This leads to significant cancellation
of terms, and one finds the acceptance probability to be \cite{Mamonov2009}

\begin{equation}
p_{\mathrm{{acc}}}(o\rightarrow n)=\min\left[1,\exp(-\beta\Delta U^{\mathrm{{rest}}})\right]\label{eq:broken hasting ratio}\end{equation}

We will also employ a second type of swap move based on \textquotedblleft{}neighbor
lists.\textquotedblright{} In the context of LBMC a neighbor list
is, for each configuration, the list of configurations deemed to be
similar by an arbitrary criterion. The trial move of interest, then,
is to choose a library configuration for swapping only among the neighboring
configurations for a single fragment $i$. When trial configurations
are selected uniformly among neighbors, it can be shown that the acceptance
criterion is \cite{Mamonov2009}

\begin{equation}
p_{\mathrm{{acc}}}(o\rightarrow n)=\min\left[1,\exp(-\beta\Delta U^{\mathrm{{rest}}})\frac{k_{i}^{o}}{k_{i}^{n}}\right]\label{eq:neighbor list acceptance ratio}\end{equation}
where $k_{i}^{o}$ and $k_{i}^{n}$ are the number of neighbors of
configuration $o$ and $n$ respectively for fragment $i$. If neighbor
lists are constructed to have the same number of neighbors for all
configurations in a given library, then the acceptance criterion of
Eq. \ref{eq:neighbor list acceptance ratio} reduces to Eq. \ref{eq:broken hasting ratio}.

Below we will explain our procedures for generating libraries and
neighbor lists.

\subsection{Practical library generation}

The fragment used in this study correspond to individual amino acids,
which are the natural building blocks of peptides. In previous LBMC
work \cite{Mamonov2009}, we used both peptide planes and amino acids
as fragments in separate simulations. However, amino acid fragments
have the advantage of including detailed \textquotedblleft{}Ramachandran
correlations\textquotedblright{} among $\phi$ and $\psi$ angles,
as well as the other degrees of freedom. In practical terms, this
means that the timescales and correlations associated with Ramachandran
effects are pre-sampled within the libraries.

Fragment configurations in the libraries were generated according
to the Boltzmann factor of OPLSAA forcefield \cite{Jorgensen1996},
plus the appropriate implicit solvent model, for all interactions
internal to the fragment. Fragment libraries must include not only
atomic coordinates, but also the six degrees of freedom necessary
for connecting one fragment to the next. Full details of the degrees
of freedom for amino acid libraries were given in \cite{Zhang2009},
our previous work. In brief, we used dummy atoms \textquotedblleft{}borrowed\textquotedblright{}
from the next fragment to facilitate sampling the coordinates necessary
for connecting fragments. Interactions with dummy atoms were fully
accounted for to yield the correct ensemble of the whole molecule
\textendash{} as can be seen in our results below.

Although library ensembles, in principle, can be generated using any
canonical sampling scheme, we found it most convenient to use internal-coordinate
Monte Carlo (ICMC). ICMC readily permits fixing degrees associated
with the dummy atoms which we did not wish to vary. Our use of ICMC
properly accounted for internal-coordinate Jacobians, which ensure
that the distribution obtained agrees with that from using the (natural)
Cartesian coordinates. The standard Jacobians were employed \---
i.e., $r^{2}$ for bond lengths $r$ and $\sin\theta$ for bond angles
$\theta$.

\begin{figure}[H]
\includegraphics[width=12cm,height=12cm]{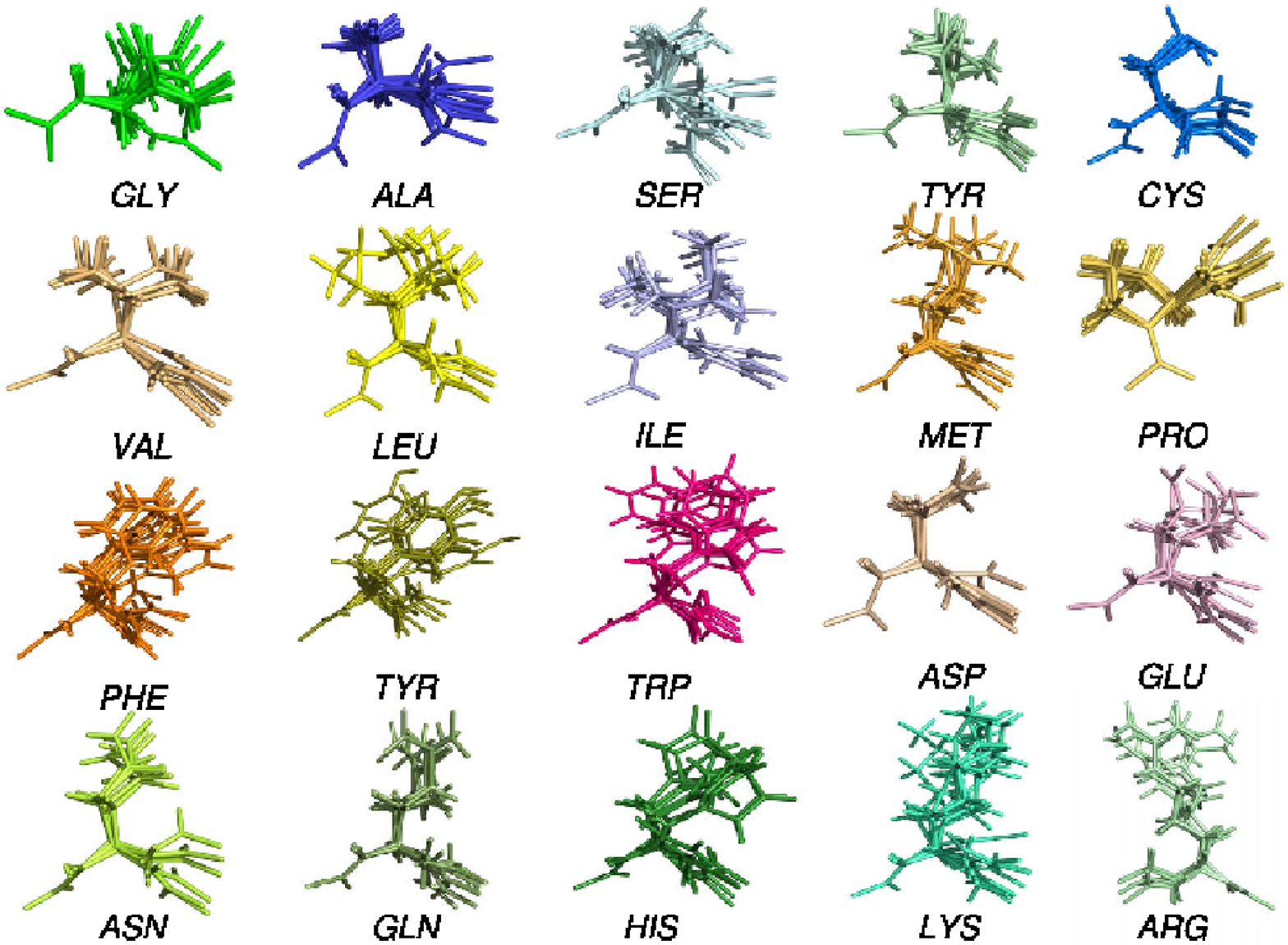}

\caption{The residue-based fragment libraries employed for library-based Monte
Carlo (LBMC) are illustrated for all 20 amino acids. We note that
the number of configurations shown in the graph does not represent
the actual library size.\label{fig:library-illustration}}

\end{figure}

For each amino acid fragment, ICMC was run for $10^{9}$ steps to
produce libraries of $10^{5}$ configurations. See Figure \ref{fig:library-illustration}.
The library configurations may not be fully statistically independent,
but we do carefully assess the statistical quality of the ultimate
ensembles of the full molecule \textendash{} as shown below.

\subsection{Neighbor-list construction}

In LBMC, {}``neighbor lists'' of library configurations similar
to each library member provide a conveninent way to attempt relatively
local moves in configuration space. As explained in our initial derivation
\cite{Mamonov2009}, neighbors can be defined in an arbitrary way.
Natural choices include criteria based on a pairwise {}``distance''
similarity metric such as the root mean square deviation (RMSD) or
the sum of absolute differences over all bond and dihedral angles
in a given fragment as was used in our previous work \cite{Mamonov2009}.
When constructing neighbor lists, if configuration $i$ contains configuration
$j$ in its neighbor list then $j$ must have $i$, to satisfy microscopic
reversibility.

In the present work the neighbor lists were constructed to generate
groups of $n=10$ similar configurations. To minimally perturb the
molecule's overall structure, similarity between two configs was quantified
by the RMSD of six atoms, three at each end of the fragment. To construct
neighbor lists, the following algorithm was used. A first {}``reference''
structure is selected randomly from the whole library and the nearest
$n-1$ configurations in the RMSD space are chosen for the first neighbor
list. The next reference structure is randomly selected from the remaining
configurations and again the closest $n-1$ configurations are chosen
for this neighbor list. This process is repeated until the whole library
is partitioned into equi-sized neighbor lists. In this study, each
library has 10$^{5}$ configurations and is partitioned into 10$^{4}$
neighbor groups of size $n=10$. 

In general, it is not necessary to make equal size clusters, nor is
it necessary to strictly partition the whole library into {}``disconnected''
neighbor lists. If there is a strict partitioning (as in the present
study), then non-neighbor trial moves are required to ensure the possibility
of ergodicity. By adjusting the fraction of local to global moves,
the acceptance rate can be tuned. In the future, it will be worthwhile
to construct and test overlapping (non-isolated) neighbor lists.

\subsection{Efficiency analysis}

It is critical to quantify the sampling quality of any new method,
in comparison to a standard technique. In this study we assess the
convergence of LBMC simulations and compare its efficiency relative
to standard Langevin dynamics using three different statistical tools.
One of these methods is semi-qualitative and the other two are quantitative. 

Because there are no true physical timescales in our Monte Carlo simulations,
our primary focus is to compare sampling efficiency in terms of single-processor
wall-clock time. We recognize that different Langevin implementations
(i.e., in different software packages) will vary in speed. However,
we anticipate such differences will be small compared to the orders
of magnitude efficiency we report below for LBMC. Furthermore, our
reference Langevin simulations employ a low friction constant, which
is recognized to improve sampling speed compared to a more physical
water-like value \cite{Paterlini1998}.

The semi-qualitative tool we use to analyze sampling is the standard
autocorrelation function of some slowly changing variable. The autocorrelation
function is given, as usual, by

\begin{equation}
C_{x}(\tau)=\frac{\langle x(t)\; x(t+\tau)\rangle-\langle x\rangle^{2}}{\langle x^{2}\rangle-\langle x\rangle^{2}}\label{eq:correlation function definition}\end{equation}
where $\langle x\rangle$ is the average value of $x(t)$, and $\tau$
is the time interval or number of MC steps between configurations
in the trajectory. Because all correlations in an LBMC {}``trajectory''
are sequential, a {}``time'' correlation description is valid. A
number of useful slow coordinates can be defined \cite{Vitalis2007},
and we choose the end-to-end distance of a peptide as a simple geometric
measure which illustrates the key timescales. However, the auto-correlation
behavior is not used to quantify efficiency in our study, but only
to depict it graphically. We measured time in units of wall-clock
minutes to facilitate comparison between LBMC and standard Langevin
simulation.

The second statistical tool is based on our previously developed {}``structural
de-correlation time'' analysis which determines how much simulation
time must elapse between configurations in the trajectory in order
for them to exhibit statistical independence \cite{Lyman2007}. The
ratio of the overall trajectory length to the decorrelation time provides
an objective estimate of the effective sample size (ESS) \--- i.e.,
the number of independent configurations. Importantly, because all
correlations are sequential in the LBMC Markov chain, the ESS for
LBMC can be calculated from the same ratio of trajectory length to
decorrelation time.

We therefore define the first efficiency factor as the gain in the
sampling speed of LBMC over Langevin dynamics based on the ratio of
CPU cost per independent configuration:

\begin{equation}
\hat{\gamma}_{1}=\frac{ESS_{\mathrm{\mathrm{LBMC}}}/t_{\mathrm{\mathrm{LBMC}}}}{ESS_{\mathrm{Langevin}}/t_{\mathrm{Langevin}}}\label{eq:efficiency definition1}\end{equation}
where $t_{\mathrm{LBMC}}$ and $t_{\mathrm{Langevin}}$ are the total
wallclock times of LBMC and Langevin simulation respectively. 

The last statistical method is based on the more traditional block
averaging analysis \cite{Flyvbjerg1989,Grossfield2009} of some slowly
changing variable. In this approach, a trajectory is divided into
{}``blocks'' of different size. The mean value of the variable along
with the standard error of the mean ($SE$) is calculated for different
size blocks. As the block size increases, so does the standard error
because blocks become more independent from each other. At some block
size the standard error levels off, indicating that the blocks have
become effectively independent from each other. This plateau is the
true value of $SE$. We use block-averaging of the end-to-end distance,
as a representative slow coordinate. We therefore define the second
efficiency factor based on the ratio of CPU cost per {}``unit of
precision'':

\begin{equation}
\hat{\gamma}_{2}=\frac{t_{\mathrm{Langevin}}SE_{\mathrm{Langevin}}^{2}}{t_{\mathrm{LBMC}}SE_{\mathrm{LBMC}}^{2}}\label{eq:efficiency definition2}\end{equation}
where $SE_{\mathrm{Langevin}}$ and $SE_{\mathrm{LBMC}}$ are the
standard errors for Langevin and LBMC simulation, respectively, estimated
from block averaging. Note that $SE^{2}$ is expected to vary linearly
with inverse simulation time \cite{Grossfield2009}.

\subsection{System and simulation details}

We applied LBMC to three implicitly solvated peptides including Ace-(Ala)$_{\text{4}}$-Nme,
Ace-(Ala)$_{\text{8}}$-Nme, and Met-enkephalin described by OPLS-AA
forcefield \cite{Jorgensen1996}. No cutoffs were used in these relatively
small systems. We chose these peptides because they have been extensively
studied experimentally and computationally \cite{Hansmann1997,Hudgins1999,Schweitzer-Stenner2004}.
Two different implicit solvent models were employed: a uniform dielectric
constant of $\varepsilon=60$ and the more standard GBSA model \cite{Qiu1997}.
The constituent atomic radii for GBSA are taken from the OPLS-AA force
field and the nonpolar solvation is calculated via the ACE approximation
\cite{Schaefer1996}. The Born radii used in GBSA are recomputed for
every MC step. 

For LBMC simulations of poly-alanine systems, three libraries were
employed corresponding to Ace, Ala and Nme fragments. For Met-enkephalin
(Tyr-Gly-Gly-Phe-Met), six libraries were used corresponding to Ace,
Gly, Phe, Tyr, Met and Nme residues. Different libraries were used
depending on the solvent models. For LBMC with uniform dielectric
solvent we used fragment libraries sampled according to the uniform
dielectric model, whereas for GBSA simulations we used libraries sampled
according to GBSA solvent. 

For all LBMC simulations reported here, the trial move was a single
fragment swap with the corresponding library. For our systems, this
was found to be the most efficient based on simulations with different
number of fragments swapped per MC step. All system were sampled at
298 K.

For LBMC simulations of both solvent models Ace-(Ala)$_{\text{4}}$-Nme
was run for 10$^{\text{5}}$ MC steps with configurations saved every
10 MC steps resulting into 10$^{\text{4}}$ frames. Ace-(Ala)$_{\text{8}}$-Nme
was run for 10$^{7}$ MC steps with frames saved every 100 MC steps
resulting in 10$^{\text{5}}$ structures. Met-enkephalin was run for
10$^{6}$ MC steps with frames saved every 10 MC steps resulting into
10$^{\text{5}}$ frames.

To compare LBMC with Langevin dynamics we ran LD simulations for the
same three systems and both solvent models. Specifically, all three
peptides were run for 100 ns with frames saved every picosecond resulting
into 10$^{\text{5}}$ structures. All Langevin simulations were run
at the temperature of 298 K and the friction constant of 5 psec$^{\text{-1}}$,
as implemented in the Tinker software package \cite{Ponder1987}.

\section{Results}

\subsection{Ensemble Quality}

\begin{figure}[H]
A

\includegraphics[width=12cm,height=5.5cm]{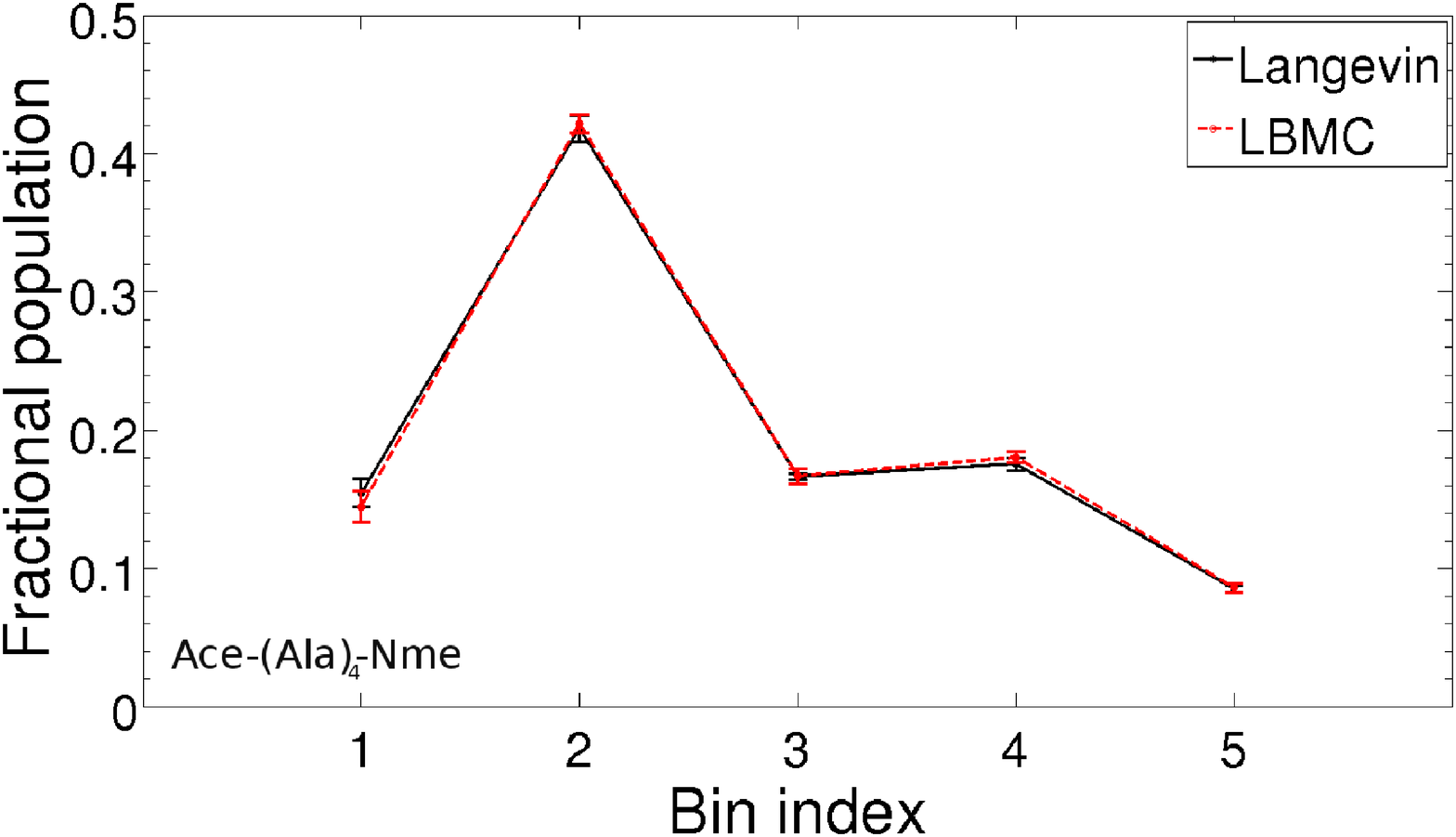}

B

\includegraphics[width=12cm,height=5.5cm]{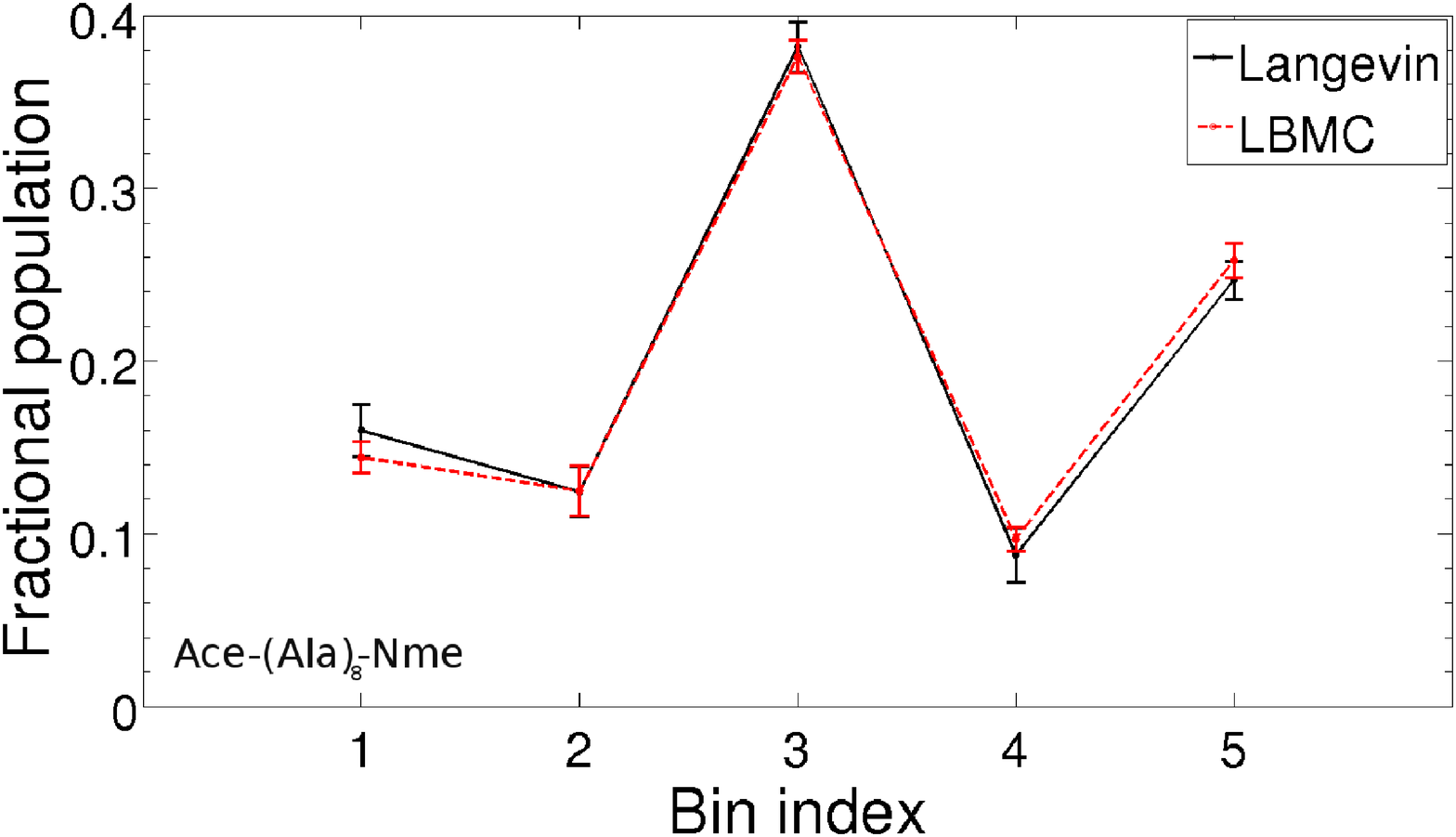}

C

\includegraphics[width=12cm,height=5.5cm]{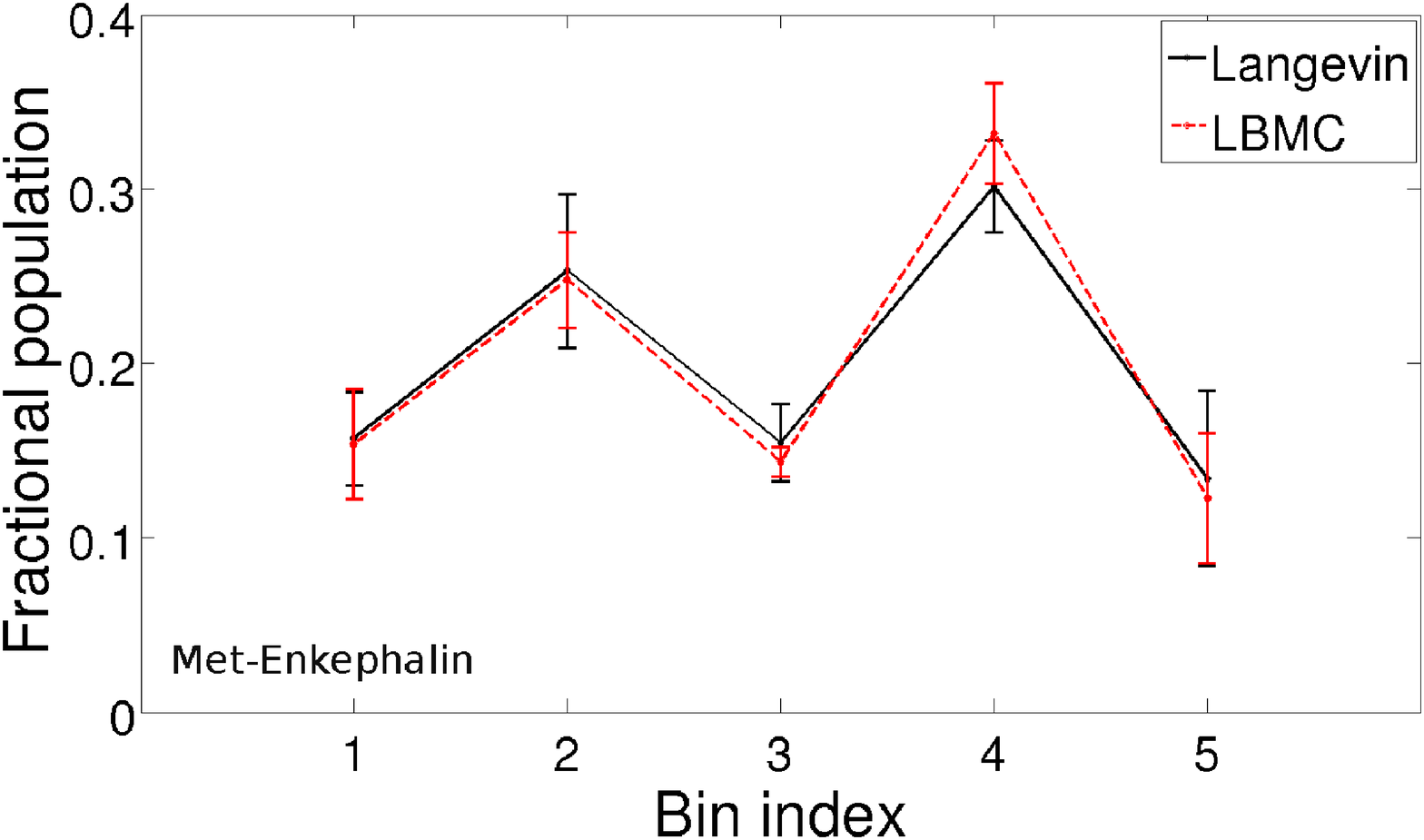}

\caption{Confirmation of correct equilibrium sampling in simple solvent. Fractional
population of structural bins obtained from LBMC (red dashed line)
and Langevin simulations (black solid line) are shown for three peptides:
(A) Ace-(Ala)$_{\text{4}}$-Nme, (B) Ace-(Ala)$_{\text{8}}$-Nme and
(C) Met-enkephalin. The peptides were sampled according to the OPLS-AA
forcefield with the uniform dielectric of 60 to model the solvent.
Error bars represent one standard deviation for each bin, calculated
from 10 independent simulations for both LBMC and Langevin.\label{fig:distribution plot eps 60}}

\end{figure}

\begin{figure}[H]
A

\includegraphics[width=12cm,height=5.5cm]{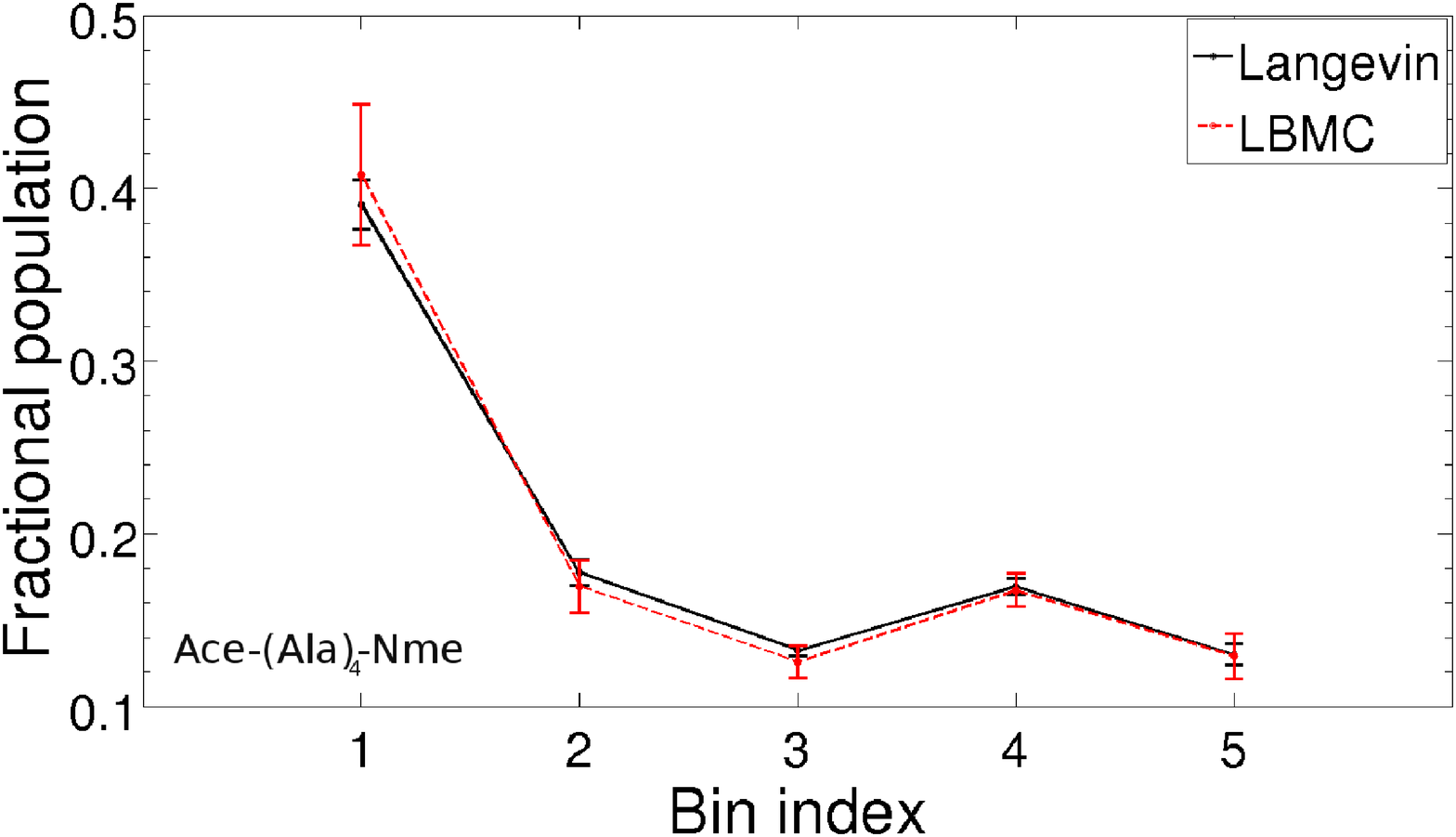}

B

\includegraphics[width=12cm,height=5.5cm]{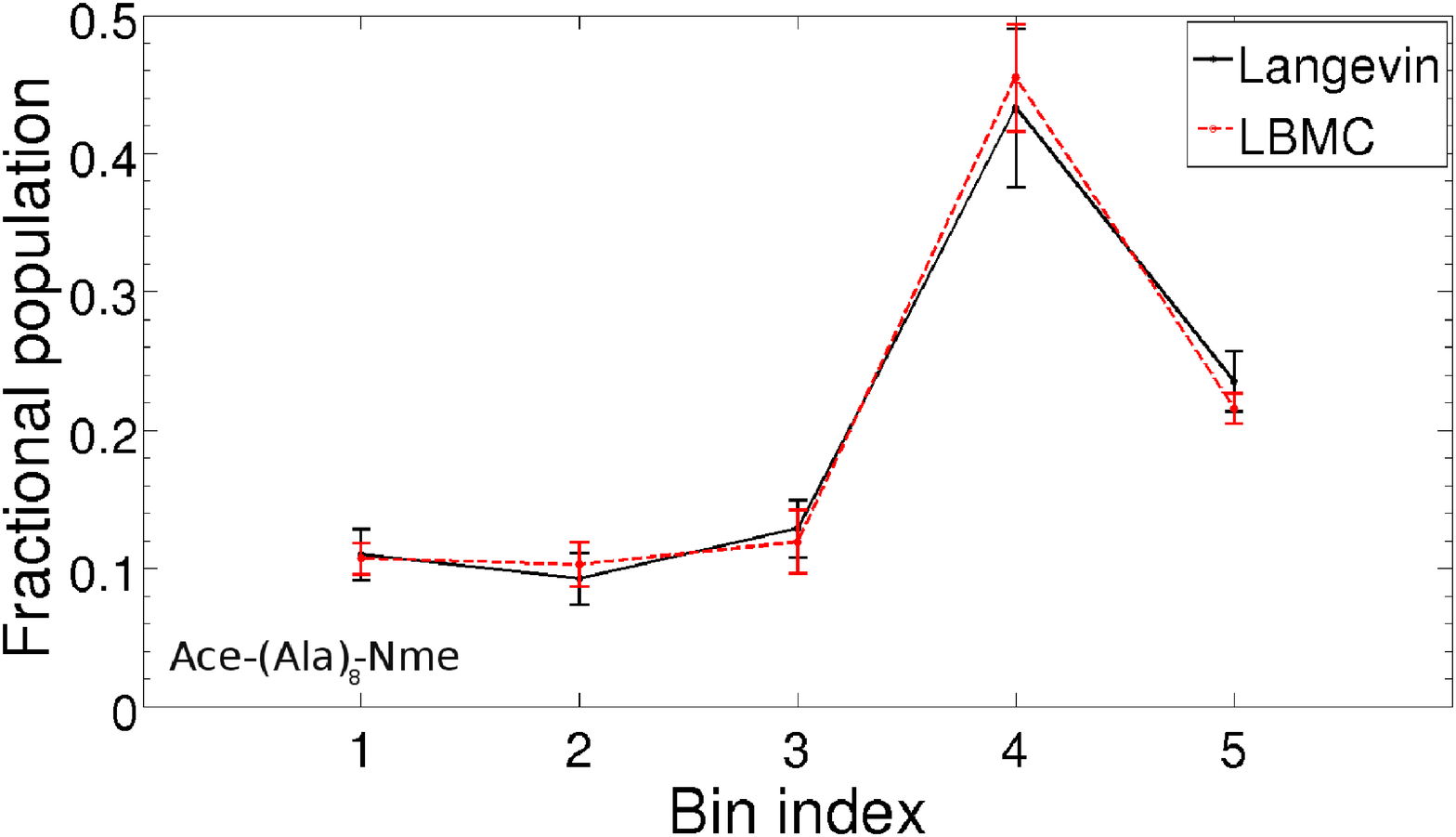}

C

\includegraphics[width=12cm,height=5.5cm]{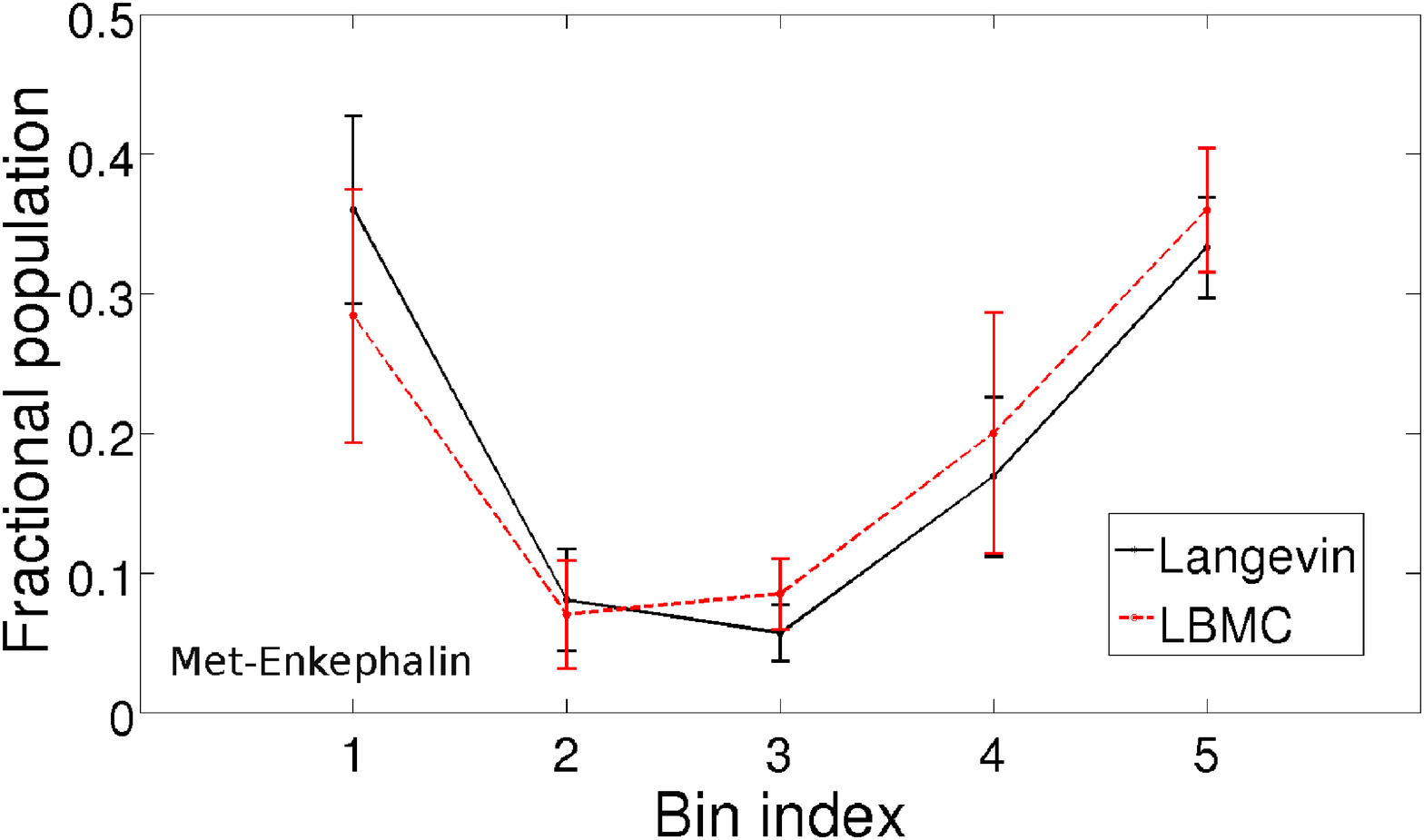}

\caption{Confirmation of correct equilibrium sampling in GBSA. Fractional population
of structural bins obtained from LBMC (red dashed line) and Langevin
simulations (black solid line) are shown for three peptides: (A) Ace-(Ala)$_{\text{4}}$-Nme,
(B) Ace-(Ala)$_{\text{8}}$-Nme (C) Met-enkephalin. The peptides were
sampled according to the OPLS-AA forcefield with GBSA solvent. Error
bars represent one standard deviation for each bin, calculated from
10 independent simulations of both LBMC and Langevin.\label{fig:distribution plot gbsa}}

\end{figure}

We first verified that LBMC samples the correct equilibrium distributions.
For this purpose we prepared {}``structural bins'' which are randomly
selected regions of configuration space which cover the whole space,
and can sensitively quantify sampling \cite{Lyman2006}. The bins
were constructed using a Voronoi procedure as described in \cite{Zhang2009}.
For all three systems we compared the fractional populations of the
structural bins obtained from LBMC and Langevin simulations. The results
for the uniform dielectric solvent model are shown in Figure \ref{fig:distribution plot eps 60}
and for GBSA in Figure \ref{fig:distribution plot gbsa}, indicating
good agreement between two methods.

To examine the ensembles from a more traditional perspective, we also
calculated hydrogen bond population and the helical content of octa-alanine.
Based on the hydrogen-bond definition given in \cite{Abrams2008},
we find the average number of hydrogen bonds in tetraalanine, octaalanine
and Met-enkephalin to be ($2.400\pm0.006$ , $5.00\pm0.02$, $3.20\pm0.01$)
in the simple solvent model and ($2.84\pm0.08$ , $6.63\pm0.06$,
$3.98\pm0.07$) in GBSA from LBMC simulation. For comparison, we found
($2.408\pm0.002$ , $5.03\pm0.02$, $3.21\pm0.01$) in simple solvent
model and ($2.75\pm0.08$ , $6.51\pm0.06$, $4.06\pm0.09$) in GBSA
from Langevin simulation. Here the uncertainty is quantified as the
standard error of mean of the number of hydrogen bonds in the specific
ensemble. Helical content was defined to be the fraction of residues
in the system whose $(\phi$, $\psi)$ dihedrals were within $\pm25^{\circ}$
of the ideal angles of approximately $(-60^{\circ},-40^{\circ})$.
We found helical population for octaalanine is  $11.7\%(\pm0.4\%)$
in the simple solvent model and $30\%(\pm2\%)$ in GBSA from LBMC
as compared $11.1\%(\pm0.3\%)$ in simple solvent model and $31\%(\pm2\%)$
in GBSA from Langevin simulation, respectively. These structural measures
further verify our results.

\subsection{Efficiency Analysis}

\begin{figure}[H]
A 

\includegraphics[width=9cm,height=5.5cm]{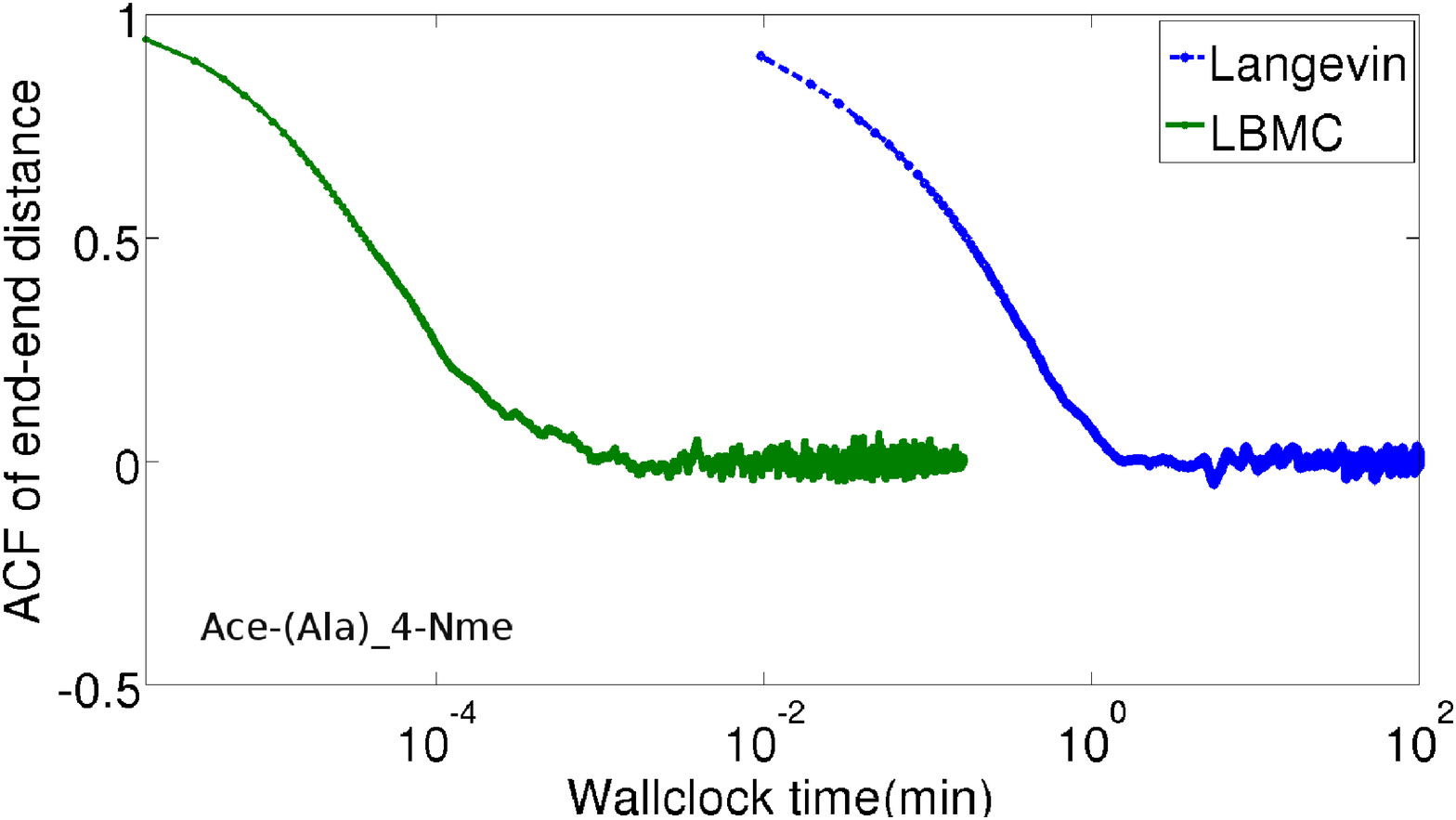}\includegraphics[width=9cm,height=5.5cm]{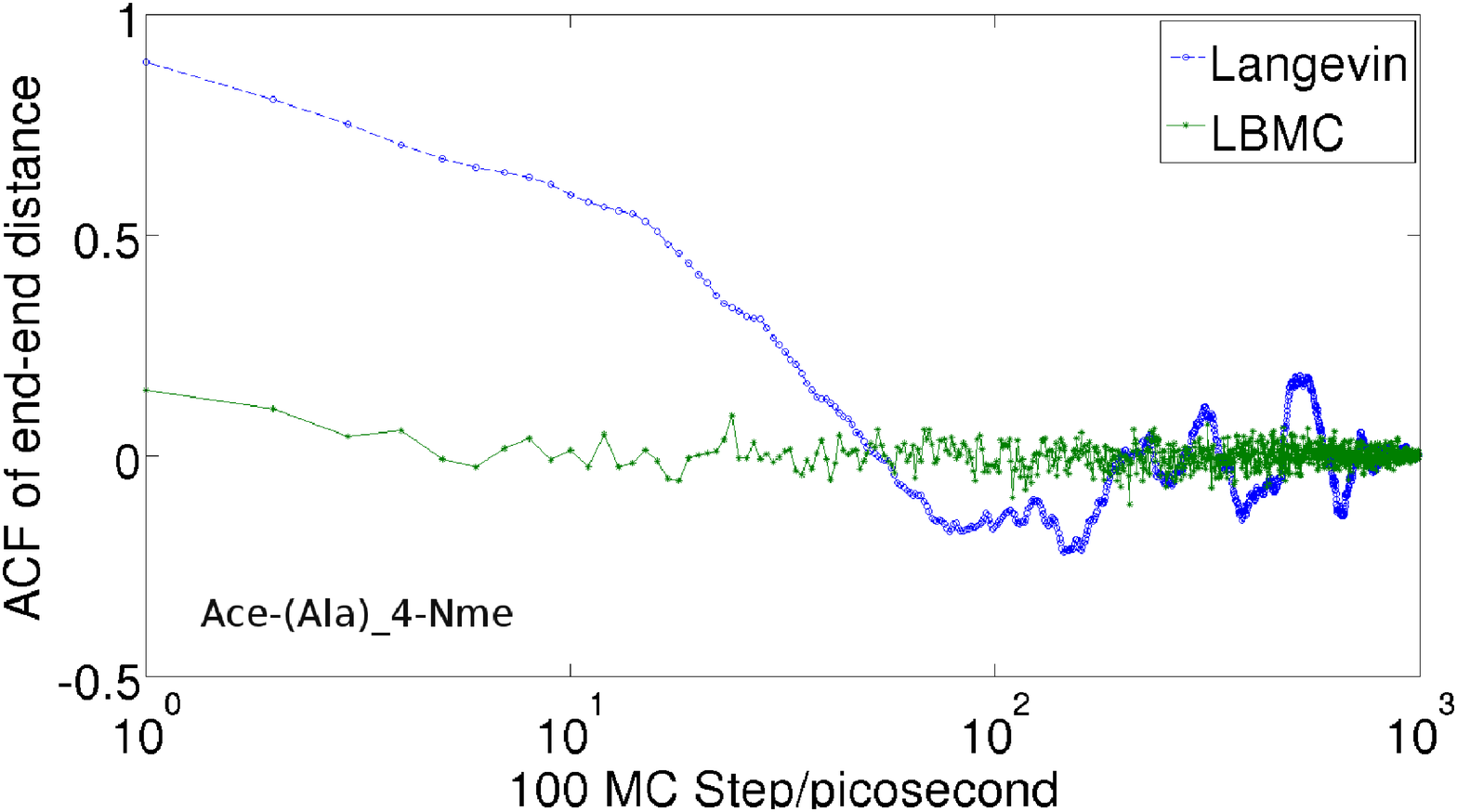}

B

\includegraphics[width=9cm,height=5.5cm]{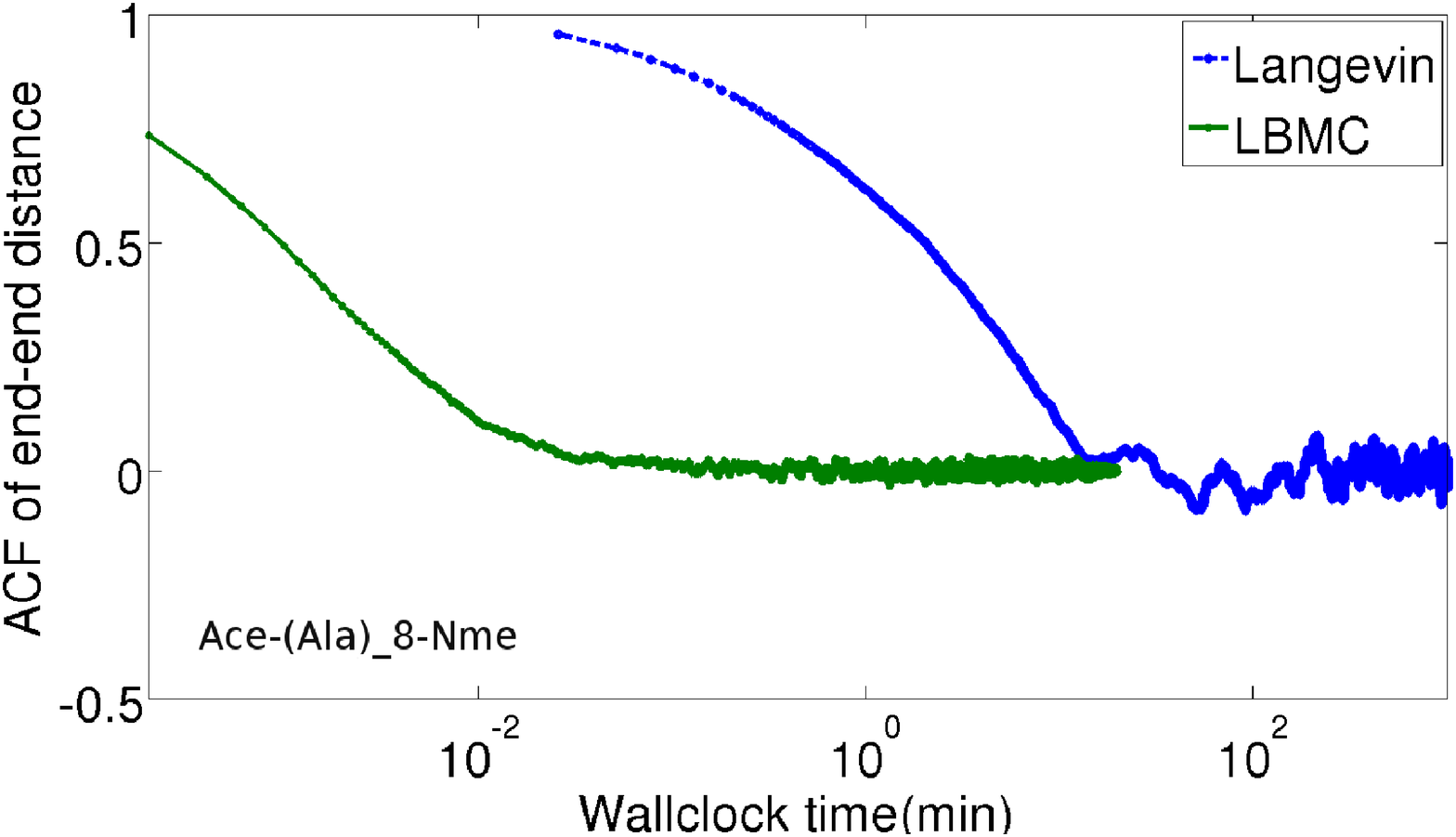}\includegraphics[width=9cm,height=5.5cm]{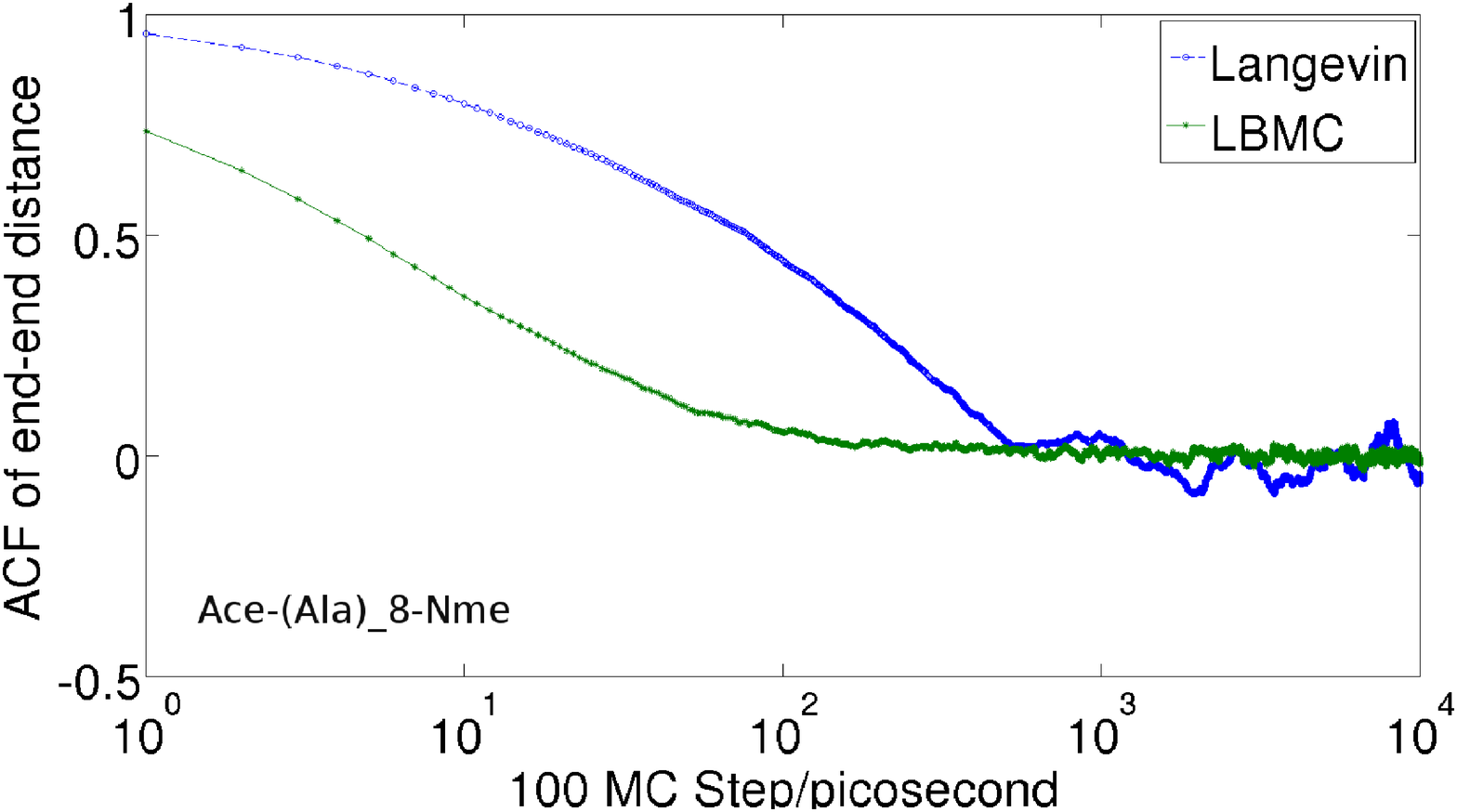}

C

\includegraphics[width=9cm,height=5.5cm]{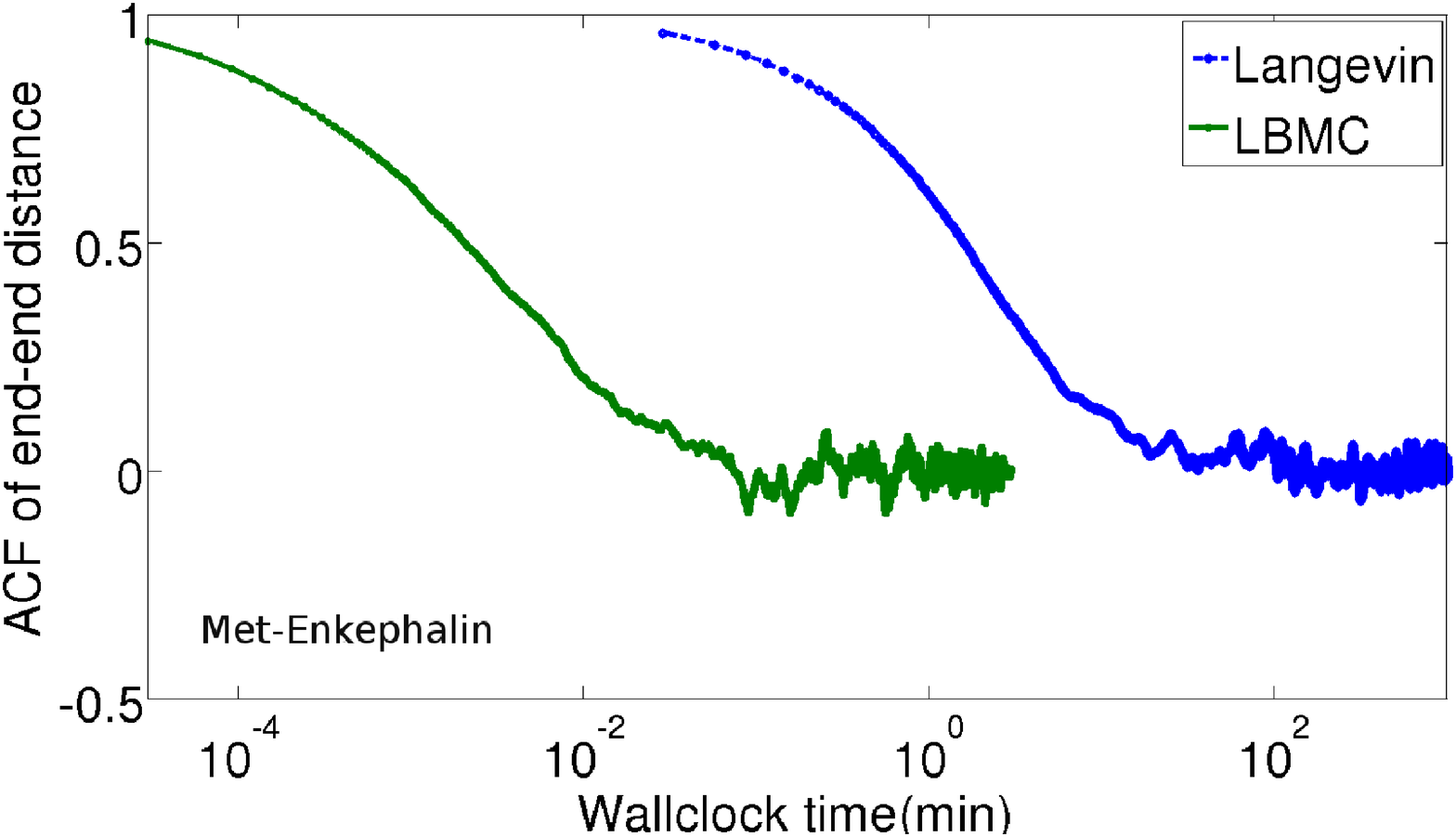}\includegraphics[width=9cm,height=5.5cm]{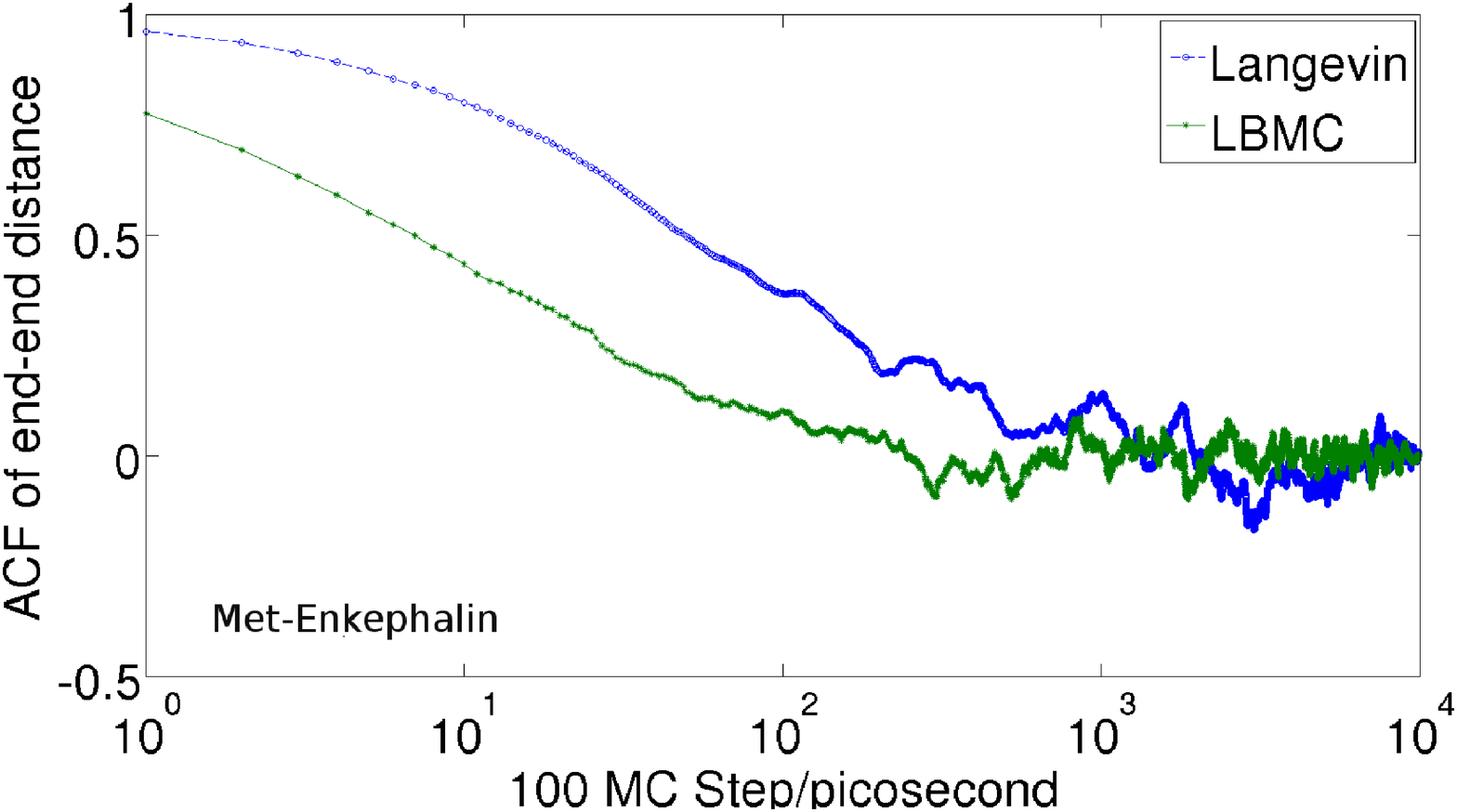}

\caption{Comparison of autocorrelation functions in simple solvent for three
peptides based on LBMC (green) and Langevin simulations (blue). The
left column shows the autocorrelation function (ACF) of the end-to-end
distance vs wallclock time and the right column shows the ACF vs timestep:
(A) Ace-(Ala)$_{\text{4}}$-Nme, (B) Ace-(Ala)$_{\text{8}}$-Nme,
(C) Met-enkephalin. The peptides were sampled according to the OPLS-AA
forcefield with a uniform dielectric of 60 to model the solvent. \label{fig:ACF under simple solvent}}

\end{figure}

\begin{figure}[H]
A

\includegraphics[width=9cm,height=5.5cm]{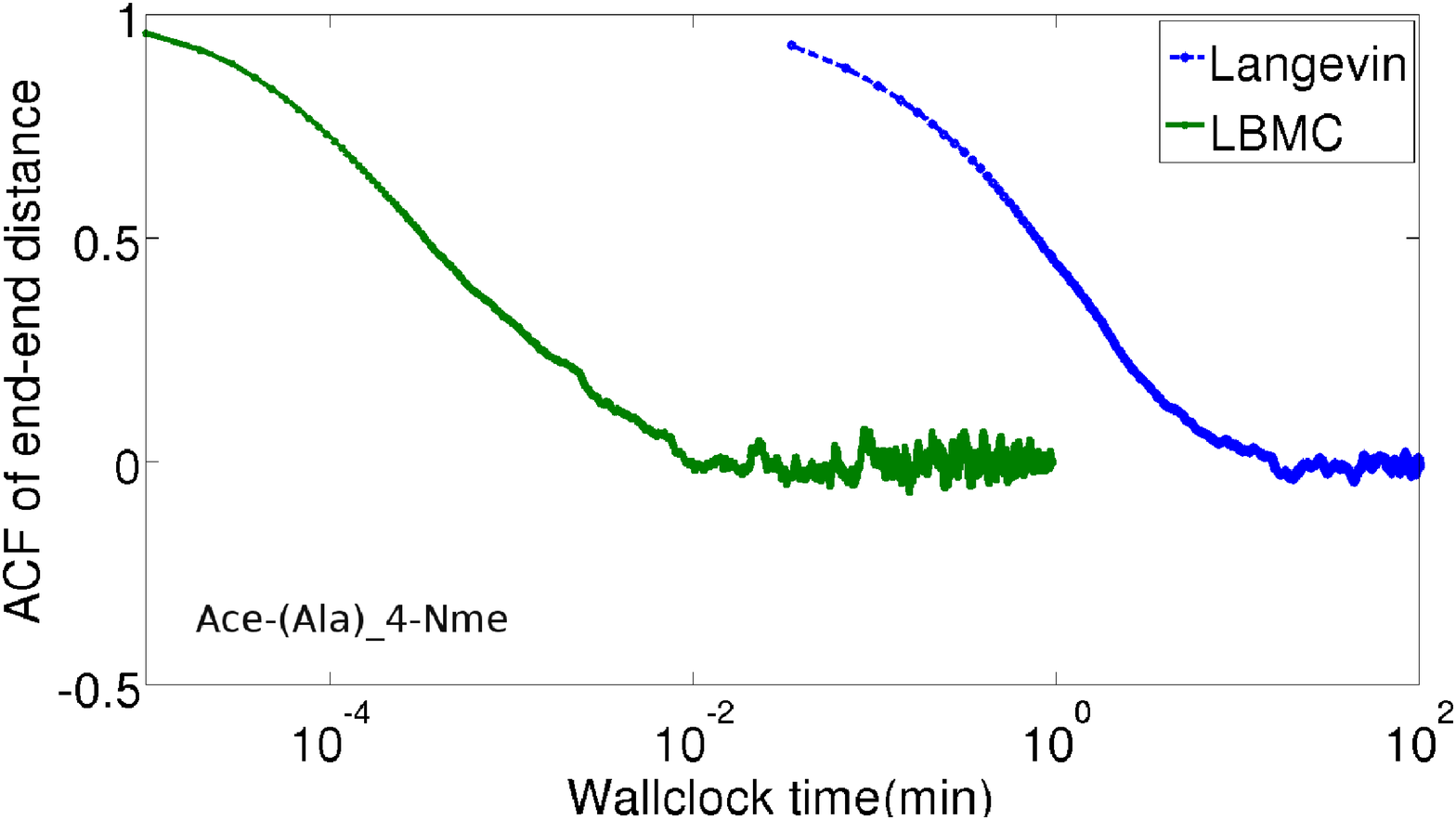}\includegraphics[width=9cm,height=5.5cm]{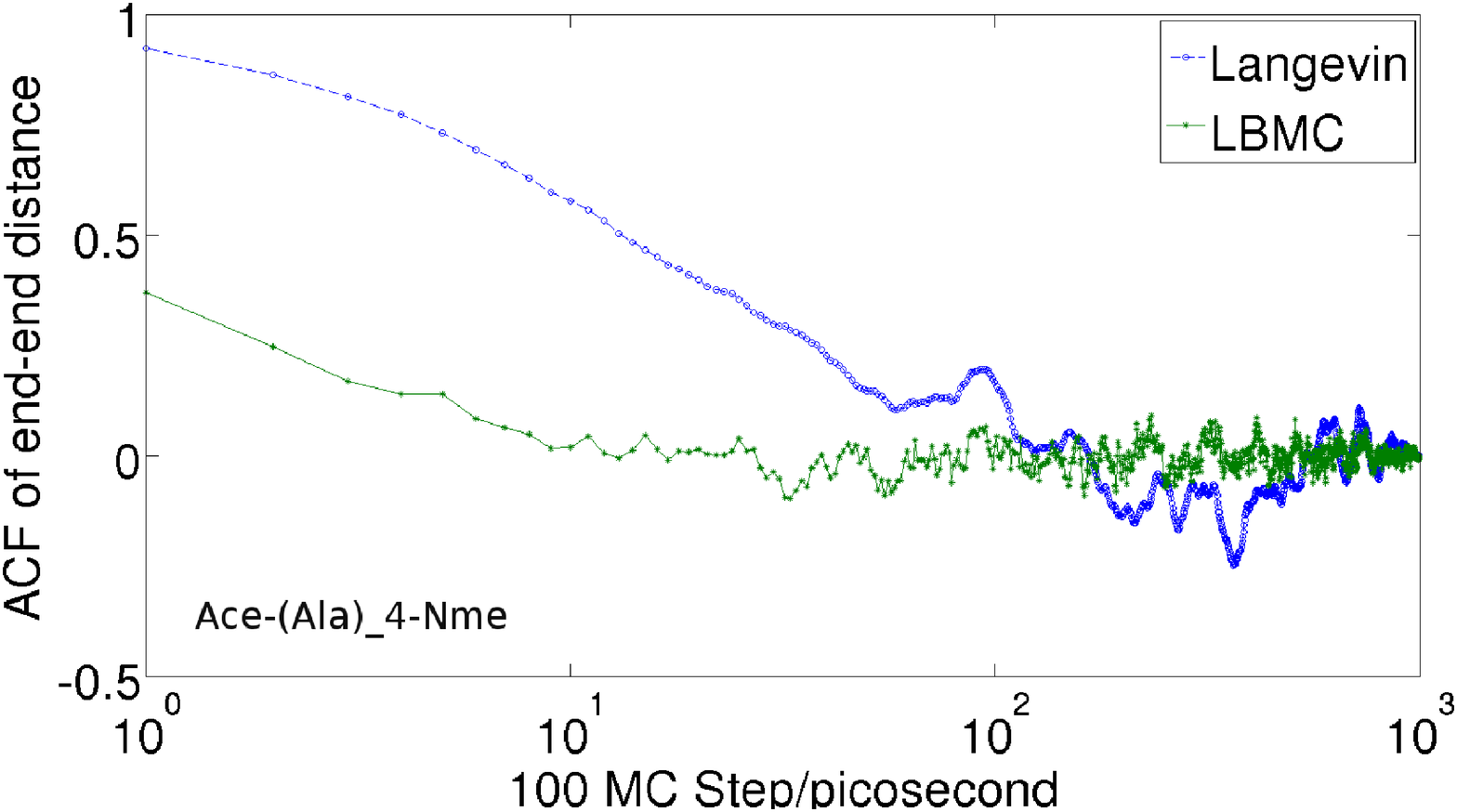}

B

\includegraphics[width=9cm,height=5.5cm]{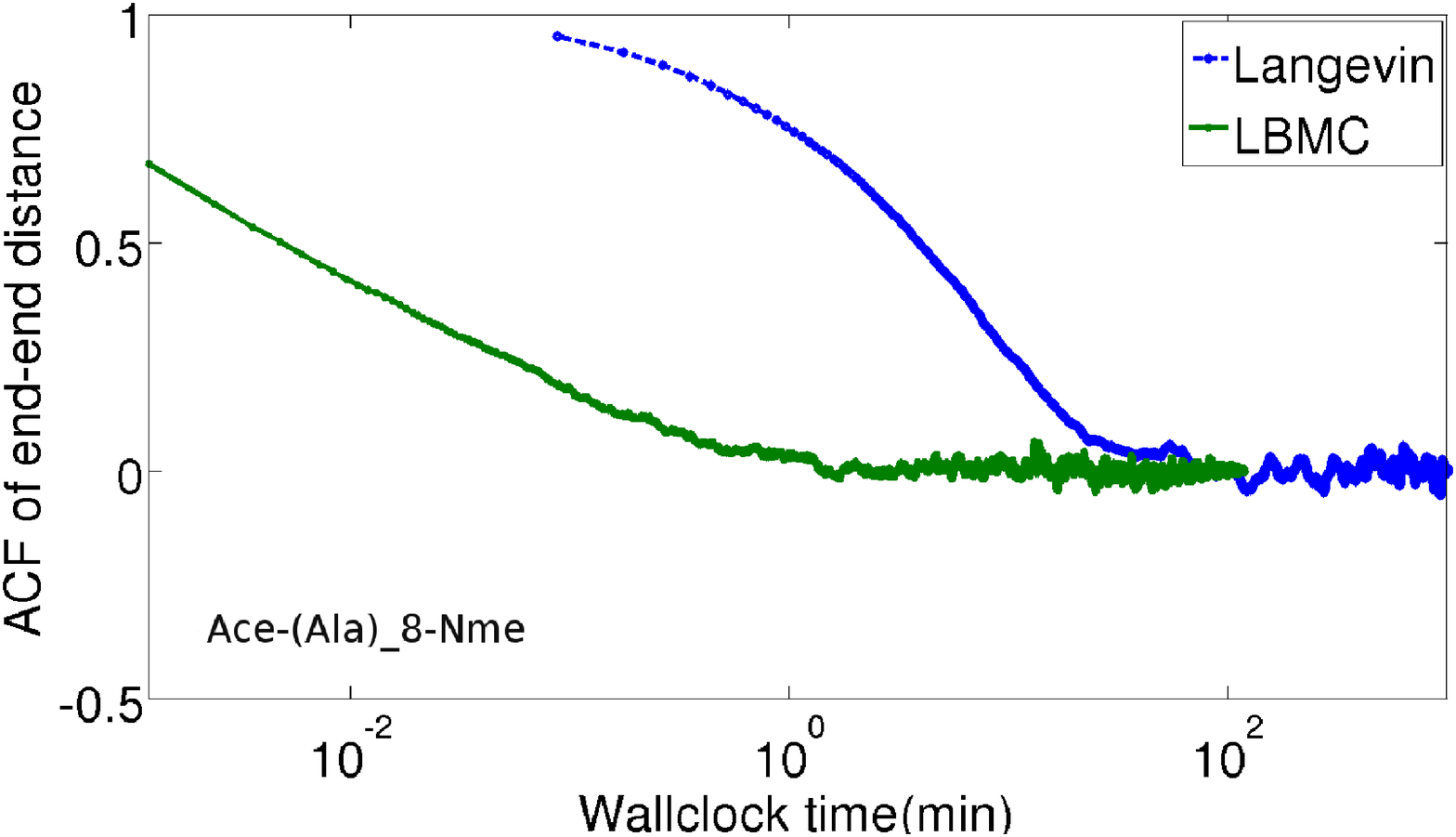}\includegraphics[width=9cm,height=5.5cm]{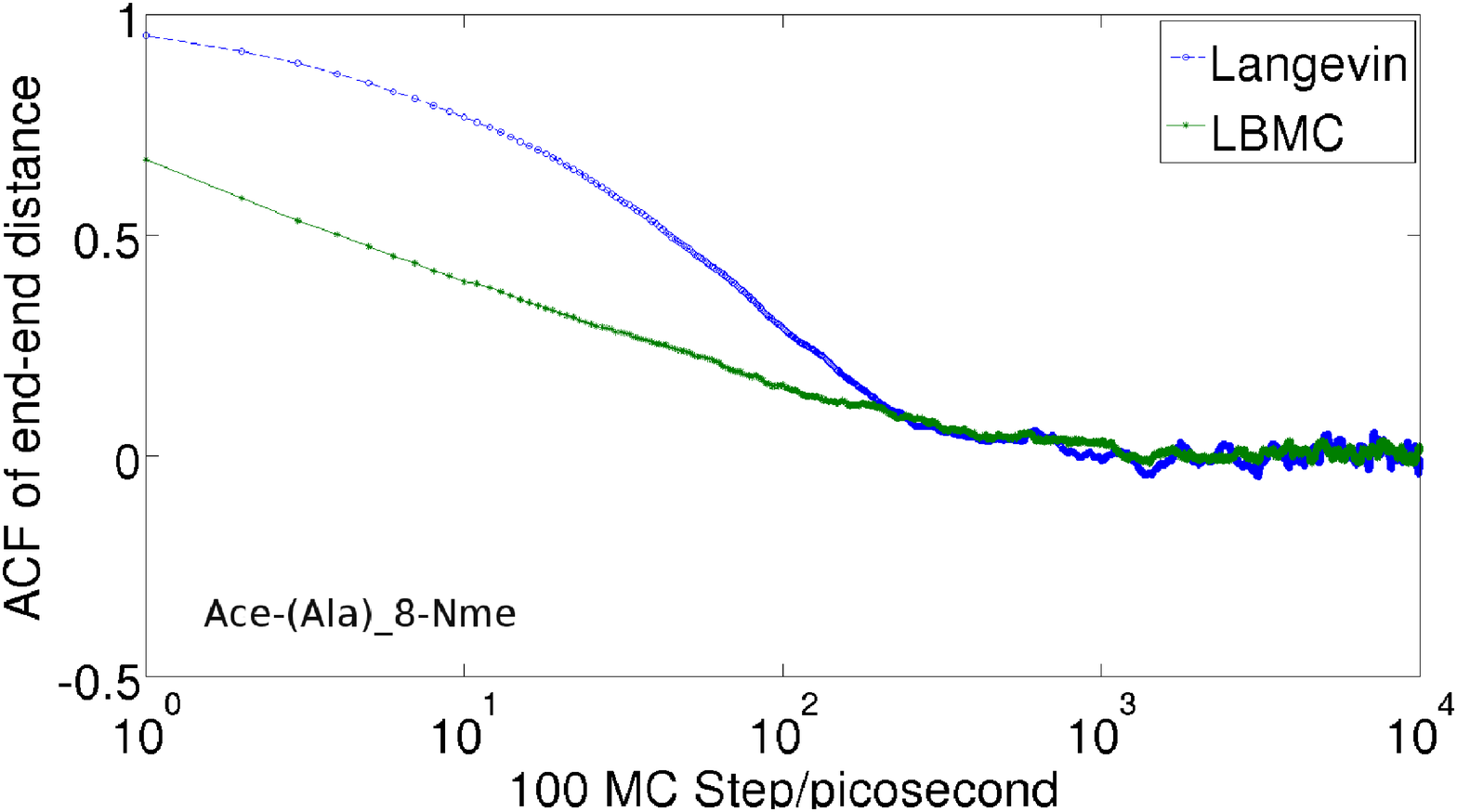}

C

\includegraphics[width=9cm,height=5.5cm]{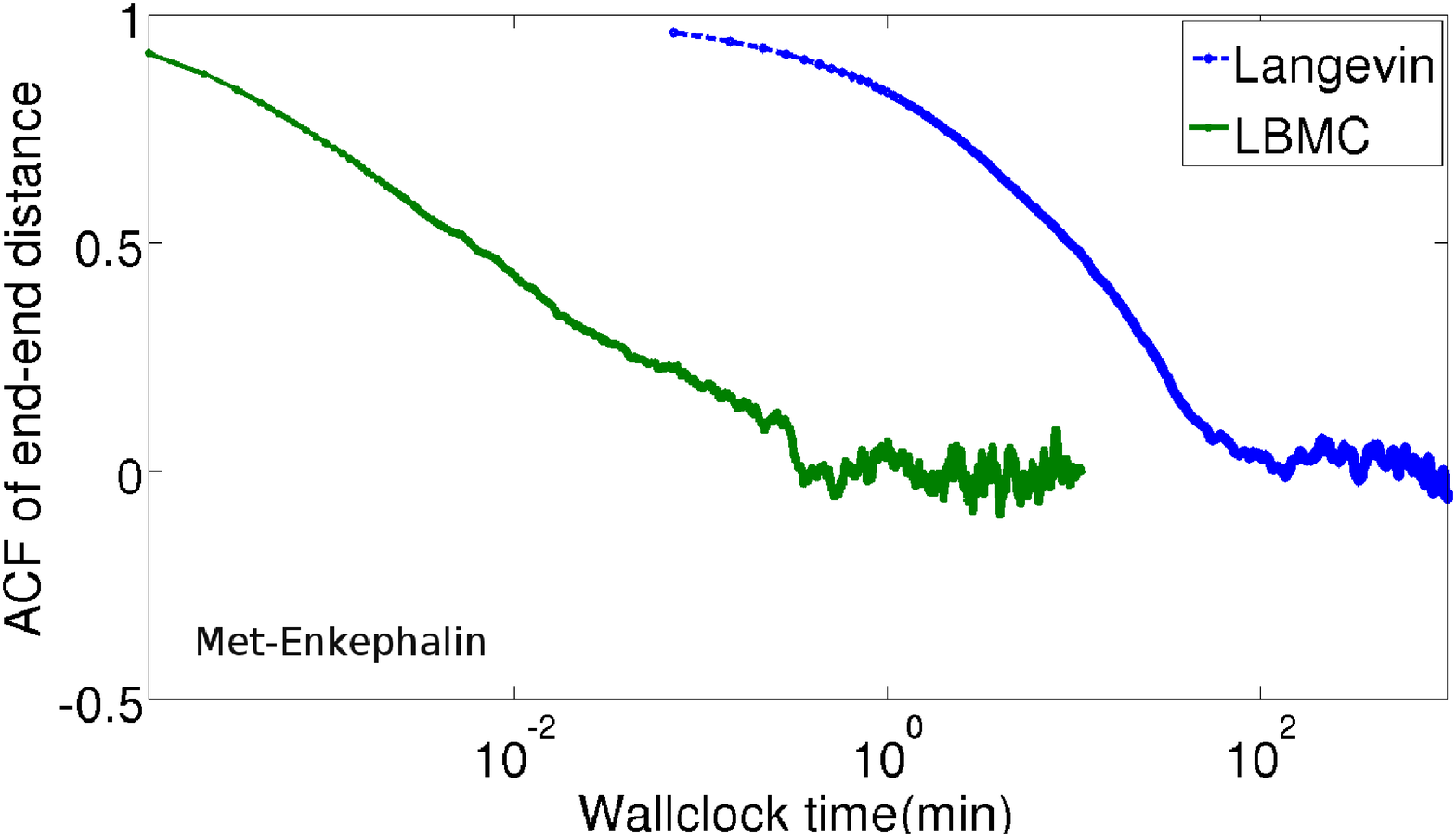}\includegraphics[width=9cm,height=5.5cm]{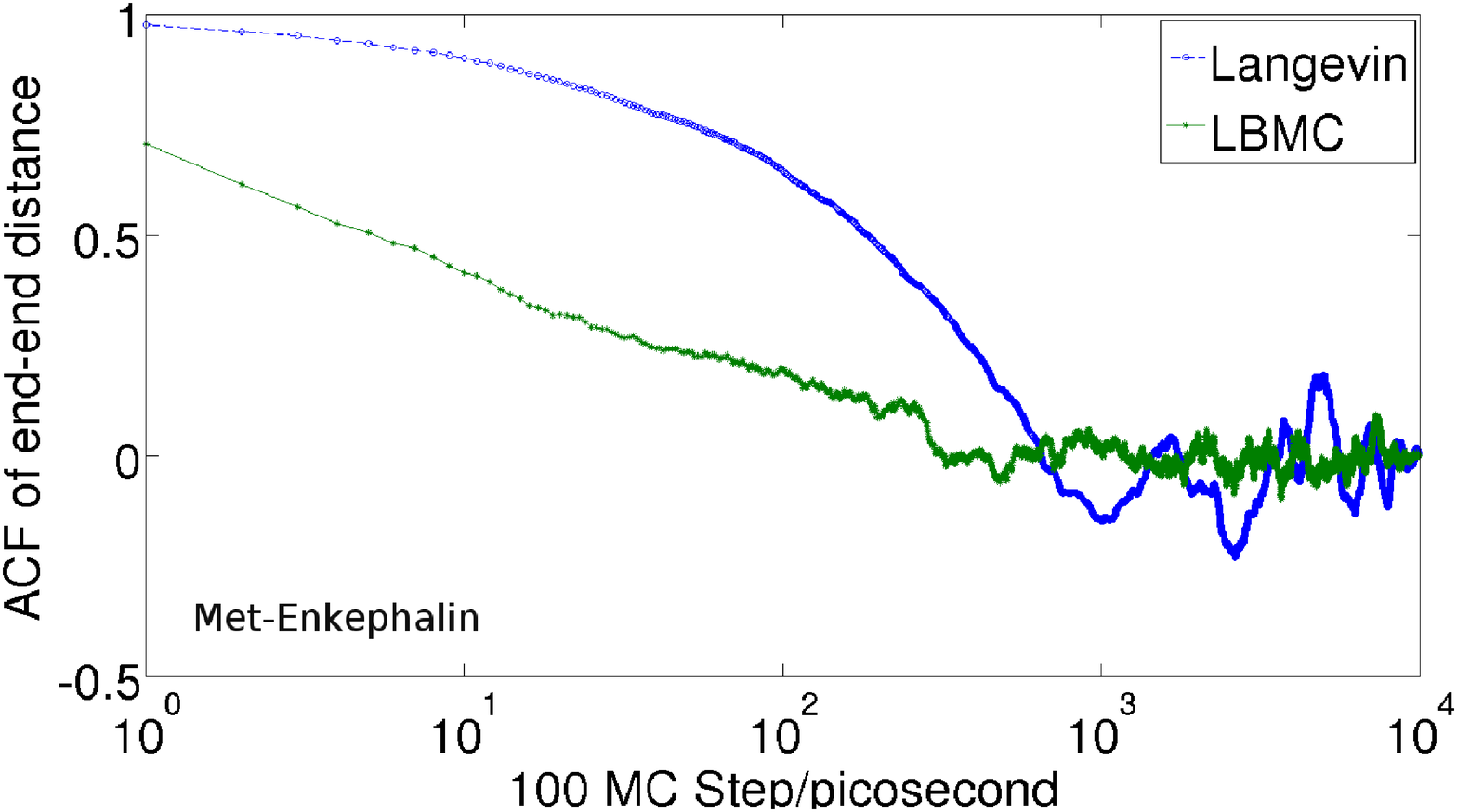}

\caption{Comparison of autocorrelation functions in GBSA for three peptides
between LBMC (green) and Langevin simulations (blue). The left column
shows the autocorrelation function (ACF) of the end-to-end distance
vs wallclock time and the right column shows the ACF vs timestep:
(A) Ace-(Ala)$_{\text{4}}$-Nme, (B) Ace-(Ala)$_{\text{8}}$-Nme,
(C) Met-enkephalin. The peptides were sampled according to the OPLS-AA
forcefield with the GBSA implicit solvent model.\label{fig:ACF under GBSA}}

\end{figure}

The efficiency of LBMC to sample the equilibrium distributions was
compared to Langevin using the three statistical tools discussed in
Sec. 2.4. The first tool is the autocorrelation function (ACF) of
the end-to-end distance for each peptide. For all systems, the end-to-end
distances was calculated based on coordinates of the methyl carbon
atom of the Ace group and the methyl carbon of the Nme group. The
autocorrelation function was calculated according to Eq. \ref{eq:correlation function definition}.
As shown in Figures \ref{fig:ACF under simple solvent} and \ref{fig:ACF under GBSA}
, we calculated ACFs for all systems in the two solvent models and
using two time measures. Most importantly, we depict the ACF vs. wallclock
time, which suggests the high efficiency of LBMC compared to Langevin
in these systems. For reference, we also computed each ACF as a function
of the number of simulation steps.

The second statistical method is the {}``de-correlation time'' analysis
which we used to calculate the number of statistically independent
configurations (i.e., the effective sample size ($ESS$)) in the trajectory
\cite{Lyman2007}. The $ESS$ results for LBMC and Langevin simulations,
along with the efficiency factors $\hat{\gamma}_{1}$ calculated using
Eq. \ref{eq:efficiency definition1}, are reported for the uniform
dielectric model in Table \ref{tab:efficiency comparison eps60} and
for GBSA in Table \ref{tab:efficiency comparison gbsa}. From Table
\ref{tab:efficiency comparison eps60} it follows that for the uniform
dielectric model LBMC is more than three orders of magnitude faster
than Langevin for Ace-(Ala)$_{\text{4}}$-Nme, more than two orders
of magnitude faster for Ace-(Ala)$_{\text{8}}$-Nme and almost three
orders of magnitude faster for Met-enkephalin. Table \ref{tab:efficiency comparison gbsa}
indicates that for GBSA solvent LBMC is more than three orders of
magnitude faster than Langevin for Ace-(Ala)$_{\text{4}}$-Nme, over
two order of magnitude faster for Ace-(Ala)$_{\text{8}}$-Nme and
over two orders of magnitude faster for Met-enkephalin. For Langevin
simulations, the decorrelation time is also a physical timescale \cite{Lyman2007}
as tabulated in Table \ref{tab:efficiency comparison eps60} and Table
\ref{tab:efficiency comparison gbsa}: it is about 1nsec or less in
the three peptides.

The third statistical tool is based on the more traditional block
averaging analysis and we used it to confirm the efficiency results
obtained from the previous method. We again employed the end-to-end
distance which is a slowly changing variable. The standard errors
($SE$) of the mean for end-to-end distances from the block averaging,
along with the efficiency factors $\hat{\gamma}_{2}$ calculated using
Eq. \ref{eq:efficiency definition2}, are reported for the uniform
dielectric model in Table \ref{tab:efficiency comparison eps60} and
for GBSA in Table \ref{tab:efficiency comparison gbsa}. Comparison
of $\hat{\gamma}_{1}$ with $\hat{\gamma}_{2}$ indicates that the
block averaging technique estimates similar efficiency factors to
the de-correlation analysis, demonstrating the robustness of our analysis.

\subsection{Regarding GBSA}

GBSA affects both simulation cost (per timestep) and the ability to
sample (by changing the landscape's roughness). Both these factors,
in turn, affect efficiency. The cost, however, is implementation specific.
We now briefly address GBSA efficiency and implementation. For GBSA,
the efficiency factors are slightly smaller than for the uniform dielectric
model. When using GBSA solvent the ESS decreased by the factor of
ca. 2.5 for both Langevin and LBMC. For LBMC the acceptance rate decreased
as well. This indicates that sampling becomes more difficult for both
methods in the more complicated energy landscape provided by GBSA.
Note that all other parameters, such as the number of atoms and the
number of steps, was the same for each method.

We can also compare solely the wallclock cost to run the same number
of MC or Langevin steps for simple solvent model and GBSA. When GBSA
solvent was employed, the simulation time increased by a factor of
4 for Langevin and by a factor of 6 for LBMC. The larger increase
of the wallclock time for LBMC can be attributed to a relatively inefficient
implementation of GBSA in our algorithm compared to the Tinker program.
We therefore believe that the decrease of efficiency factors, $\hat{\gamma}_{1}$
and $\hat{\gamma}_{2}$, for GBSA simulations can be attributed to
our inefficient implementation of GBSA rather than the difficulty
of LBMC to sample complex energy landscapes. See further comments
on GBSA in Sec. 4.

\subsection{Neighbor-based trial moves}

The use of neighbor swap moves in LBMC (Secs 2.1 and 2.3) is suggested
by Tables \ref{tab:efficiency comparison eps60} and \ref{tab:efficiency comparison gbsa},
because the acceptance rate significantly dropped for LBMC simulations
with GBSA solvent compared to the uniform dielectric model. To test
the ability of neighbor-list trial moves to increase the acceptance
rate for LBMC simulation of all-atom peptides, we employed two sets
of trial moves: one with 30\% and the other with 70\% of local (neighbor-list)
moves. Both sets helped to increase the acceptance rate. For example,
for Ace-(Ala)$_{\text{8}}$-Nme using 30\% local moves increased the
acceptance rate from 0.18 to 0.27 and 70\% local moves led to 0.41.
For Met-enkephalin 30\% local moves increased the acceptance rate
from 0.17 to 0.31 and 70\% local moves led to 0.38. However, the efficiency
analysis indicated that for simulations with local trial moves the
efficiency factors $\hat{\gamma}_{1}$ and $\hat{\gamma}_{2}$ turned
out to be smaller compared to simulations with regular (100\% global)
trial moves. As discussed in Sec 4, additional exploration of neighbor-list
construction is needed.

\section{Discussion}

The initial results for sampling all-atom peptides in implicit solvent
via LBMC are very encouraging. We wish to make some observations and
also point out several avenues for improvement as well as some limitations.

First of all, LBMC is not simply internal-coordinate MC (ICMC) with
{}``window dressing''. That is, using libraries is not merely a
way to employ large trial moves, such as drastic changes to $\phi$
and $\psi$ dihedrals. Indeed if large dihedral moves are used in
ICMC, the acceptance rate is over an order of magnitude smaller than
for global swap moves in LBMC. The key to LBMC success (in the systems
studied) is the correlated nature of trial moves: large $\phi$ and
$\psi$ changes are accompanied, by construction, with correlated
changes of other coordinates in the fragment (i.e., residue).

There appear to be two principal limitations for application of LBMC
to all-atom sampling with standard forcefields. These limitations
stem, in a sense, from the strength of LBMC for small flexible systems:
extremely large trial moves (not feasible or physical in dynamics)
lead to rapid sampling. For instance, LBMC will occasionally jump
from one region of the Ramachandran plane to a completely different
part. Such large moves immediately suggest that, first of all, LBMC
with large trial moves will not be suitable for explicit solvent.
Secondly, even in implicit solvent, once a single molecule becomes
large and \textquotedblleft{}dense\textquotedblright{} \textendash{}
such as a full protein \textendash{} large-scale trial moves will
again prove nearly impossible to accept.

We note that, in principle, LBMC is not limited to implicit solvent.
In an explicit solvent simulation of a peptide, for example, trial
moves for the peptides could be governed by LBMC and solvent moves
via {}``ordinary'' MC. In the LBMC acceptance criterion, $U^{rest}$
would include solvent interactions. Whether such an approach proves
practical will depend sensitively on the construction of suitably
local LBMC trial moves \--- i.e., crankshaft-like, see below. It
is also possible that future methodologies will permit the conversion
of implicit solvent ensembles to explicit solvent \cite{Bhatt2009}. 

LBMC can however employ more \textquotedblleft{}local\textquotedblright{}
trial moves, for instance based on \textquotedblleft{}neighboring\textquotedblright{}
library configurations as described above. As noted in the Sec. 3,
such local moves actually decrease sampling efficiency (despite increasing
the acceptance ratio) in the small systems studied here. In larger
systems, however, we expect neighbor-based moves to be helpful. Local
moves should also prove important in sampling loops with LBMC. In
the future, more sophisticated constructions of neighbor lists should
be possible, as compared to our fairly simple approach described in
Sec. 2.

Further improvements to GBSA-based sampling via LBMC appear to be
possible, given our inefficient implementation of GBSA as discussed
in Sec. 3. Indeed, generally speaking, GBSA is not well-suited to
Monte Carlo simulation, as previously noted \cite{Michel2006}, because
the non-pairwise energy terms depend on the entire molecule even when
only part of it is changed as in typical MC trial moves. Therefore,
other solvent models or approximations to GBSA \cite{Michel2006,Mongan2007,Vitalis2009}
should improve LBMC efficiency further in terms of wall-clock time.

Like almost any canonical sampling method, LBMC can be employed in
more sophisticated sampling strategies, such as replica, Hamiltonian,
or resolution exchange \cite{Swendsen1986,Geyer1991,Sugita2000,Lyman2006a},
as well as the related dual-chain MC approaches \cite{Iftimie2000,Gelb2003,Hetenyi2002}.
However, LBMC would appear to have a particular advantage for multi-resolution
approaches: the positions of all atoms can be stored at essentially
zero run-time cost, even if a \textquotedblleft{}coarse grained\textquotedblright{}
forcefield is employed. That is, because all degrees of freedom are
maintained, LBMC provides a natural means for casting resolution-exchange
simulation in terms of \textquotedblleft{}simple\textquotedblright{}
Hamiltonian exchange. We believe this idea warrants further investigation.

As noted in our earlier paper \cite{Mamonov2009} and echoing the
above discussion, LBMC provides a natural platform for coarse and
mixed resolution models. Most simply, all atoms can be accounted for,
with only a subset used as interaction sites in a \textquotedblleft{}coarse-grained\textquotedblright{}
model \textendash{} allowing flexibility to tune the coarse interactions.
In a \textquotedblleft{}mixed modeling\textquotedblright{} approach,
atoms in critical regions such as binding sites can retain their full
interactions, while distant residues are coarse-grained. Such a strategy
might be useful in binding-affinity or ensemble-based docking calculations.

\section{Summary and Conclusions}

Multiple statistical efficiency analyses show that library-based Monte
Carlo (LBMC) can obtain remarkable efficiency for peptide systems.
LBMC employs pre-calculated libraries of equilibrium configurations
of molecular fragments \textendash{} in this case, amino acid fragments.
We applied LBMC to three peptides (4, 5, and 8 residues) described
by a standard all-atom forcefield, OPLS-AA, with a simple dielectric
\textquotedblleft{}solvent\textquotedblright{} as well as the common
GBSA implicit solvent. In every case, two independent methods of quantifying
efficiency indicate that LBMC out-performed Langevin dynamics by two
orders of magnitude.

The success of LBMC in flexible peptides derives from several factors.
First, large trial moves \textendash{} with significant $\phi$ and
$\psi$ changes \textendash{} are frequently attempted. Second, because
the libraries are pre-sampled to include all interactions and correlations
internal to each residue, only long-range interactions present an
obstacle to accepting a trial move. Finally, once the libraries have
been generated they can be re-used repeatedly without the need to
re-calculate the tabulated energies internal to fragments.

LBMC thus appears to be promising for loop-sized peptides (\textasciitilde{}10
residues), particularly if trial moves to neighboring library configurations
can be better exploited. In addition, LBMC can readily be combined
with \textquotedblleft{}advanced\textquotedblright{} techniques such
as those based on the exchange idea \cite{Swendsen1986,Geyer1991,Sugita2000,Lyman2006a,Lyman2006b,Iftimie2000,Gelb2003,Hetenyi2002}.
The ability of LBMC simulation to store all atomic coordinates at
no run-time cost suggests it will provide an ideal platform for flexible
coarse-graining approaches based on using a subset of interaction
sites.

\subsubsection*{Acknowledgement}

The authors thank Xin Zhang, Bin Zhang, and Divesh Bhatt for helpful
discussions. Funding was provided by the NIH through Grants GM070987
and GM076569, as well as by the NSF through Grant MCB-0643456. 

\bibliographystyle{unsrt}
\addcontentsline{toc}{section}{\refname}\bibliography{LBMC}

\section*{\newpage}

\section*{Figures}

\newpage

\newpage

\newpage

\newpage

\section*{Tables}

\begin{table}[H]
\caption{Efficiency in simple solvent. The results of the {}``de-correlation''
and block averaging analyses of LBMC and Langevin simulations are
reported for three peptides: Ace-(Ala)$_{\text{4}}$-Nme, Ace-(Ala)$_{\text{8}}$-Nme,
and Met-enkephalin. The peptides were sampled according to the OPLS-AA
forcefield with the uniform dielectric of 60 to model the solvent.
$M$ is the number of atoms, $t$ is the total wallclock time, $t_{\mathrm{decorr}}$
is the decorrelation time of Langevin simulation in physical units,
$Acc$ is the average acceptance rate of LBMC simulation, $ESS$ is
the effective sample size, and $SE$ is the standard error of the
mean end-to-end distance. The factors $\hat{\gamma}_{1}$ and $\hat{\gamma}_{2}$
represent the efficiency gain of LBMC relative to Langevin dynamics
and are defined in Eq. \ref{eq:efficiency definition1} and Eq. \ref{eq:efficiency definition2}
respectively.\label{tab:efficiency comparison eps60}}

\begin{tabular}{c|c|cccc|cccc|cc}
\hline 
System &  & \multicolumn{4}{c|}{Langevin} & \multicolumn{4}{c|}{LBMC} &  & \tabularnewline
\cline{3-10} 
 & $M$ & $t$ & $t_{\mathrm{decorr}}$ & $ESS$ & $SE$ & $t$ & $Acc$ & $ESS$ & $SE$ & $\hat{\gamma}_{1}$ & $\hat{\gamma}_{2}$\tabularnewline
\hline
Ace-(Ala)$_{\text{4}}$-Nme & 52 & 16h & 0.08ns & 1333 & 0.07 & 10sec & 0.69 & 454 & 0.10 & 1961 & 2822\tabularnewline
Ace-(Ala)$_{\text{8}}$-Nme & 92 & 43.5h & 0.5ns & 200 & 0.22 & 20min & 0.28 & 200 & 0.10 & 145 & 632\tabularnewline
Met-enkephalin & 84 & 48h & 0.7ns & 142 & 0.26 & 3min & 0.30 & 142 & 0.27 & 924 & 839\tabularnewline
\end{tabular}
\end{table}

\newpage

\begin{table}[H]
\caption{Efficiency in GBSA implicit solvent. The results of the {}``de-correlation''
and the block averaging analyses of LBMC and Langevin simulations
are reported for three peptides: Ace-(Ala)$_{\text{4}}$-Nme, Ace-(Ala)$_{\text{8}}$-Nme
and Met-enkephalin. The peptides were sampled according to the OPLS-AA
forcefield with GBSA solvent. $M$ is the number of atoms, $t$ is
the total wallclock time, $t_{\mathrm{decorr}}$ is the decorrelation
time of Langevin simulation in physical units, $Acc$ is the average
acceptance rate of LBMC simulation, $ESS$ is the effective sample
size, and $SE$ is the standard error of the mean end-to-end distance.
The factors $\hat{\gamma}_{1}$ and $\hat{\gamma}_{2}$ represent
the efficiency gain of LBMC relative to Langevin dynamics and are
defined in Eq. \ref{eq:efficiency definition1} and Eq. \ref{eq:efficiency definition2}
respectively. \label{tab:efficiency comparison gbsa}}
\begin{tabular}{c|c|cccc|cccc|cc}
\hline 
System &  & \multicolumn{4}{c|}{Langevin} & \multicolumn{4}{c|}{LBMC} &  & \tabularnewline
\cline{3-10} 
 & $M$ & $t$ & $t_{\mathrm{decorr}}$ & $ESS$ & $SE$ & $t$ & $Acc$ & $ESS$ & $SE$ & $\hat{\gamma}_{1}$ & $\hat{\gamma}_{2}$\tabularnewline
\hline
Ace-(Ala)$_{\text{4}}$-Nme & 52 & 58h & 0.2ns & 500 & 0.11 & 58s & 0.44 & 200 & 0.17 & 1438 & 1507\tabularnewline
Ace-(Ala)$_{\text{8}}$-Nme & 92 & 147h & 0.9ns & 111 & 0.19 & 2h & 0.18 & 200 & 0.15 & 133 & 119\tabularnewline
Met-enkephalin & 84 & 120h & 2ns & 50 & 0.22 & 11min & 0.17 & 50 & 0.30 & 524 & 367\tabularnewline
\end{tabular}
\end{table}

\newpage

\tableofcontents{}
\end{document}